\documentclass[twocolumn,tighten,times,twocolappendix]{aastex701}

\usepackage{enumitem}
\usepackage{soul}
\usepackage{xcolor}
\usepackage{ulem}

\begin{document}

\DeclareGraphicsExtensions{.pdf,.png,.jpg}
\parskip 4mm

\newcommand{\Halpha}{{H$\alpha$}}
\newcommand{\FeIion}{{\ion{Fe}{1}}}
\newcommand{\FeIHMIline}{{\FeIion~$617.34$~nm}}
\newcommand{\FeIsixthreeooneline}{{\FeIion~$630.15$~nm}}
\newcommand{\FeIsixthreeotwoline}{{\FeIion~$630.25$~nm}}
\newcommand{\CaIIK}{{\ion{Ca}{2}~K}}
\newcommand{\CaIIH}{{\ion{Ca}{2}~H}}
\newcommand{\CaIIIR}{{\ion{Ca}{2}~$854.2$~nm}}
\newcommand{\MgIIk}{{\ion{Mg}{2}~k}}
\newcommand{\MgIItriplet}{{\ion{Mg}{2}~279.9~nm}}
\newcommand{\SiIV}{{\ion{Si}{4}~$139.3$~nm}}
\newcommand{\FeIX}{{\ion{Fe}{9}~$17.1$~nm}}
\newcommand{\FeXII}{{\ion{Fe}{12}~$19.5$~nm}}
\newcommand{\FeXV}{{\ion{Fe}{15}~$28.4$~nm}}

\newcommand{\jvec}{{\bf j}}
\newcommand{\vvec}{{\bf v}}
\newcommand{\Bvec}{{\bf B}}
\newcommand{\kms}{km~s$^{-1}\,$}
\newcommand{\pot}[2]{{#1\times 10^{#2}}}
\newcommand{\nl}{\hfill\break}
\newcommand{\LB}{{L_B^{}}}
\newcommand{\invLB}{{\LB^{-1}}}
\newcommand{\invLBinMm}{{\LB[\hbox{Mm}]^{-1}}}
\newcommand{\Bhor}{{B_{\mathrm{hor}}}}


\title{\Large{Magnetic flux cancellation in the solar atmosphere\\ through 3D realistic numerical modeling}}
\shorttitle{Magnetic flux cancellation}

\correspondingauthor{Fernando Moreno-Insertis}
\email{fmi@iac.es}

   \author{Fernando Moreno-Insertis}
   \email{fmi@iac.es}
   \affiliation{Instituto de Astrofisica de Canarias, E-38205 La Laguna, Tenerife, Spain}
   \affiliation{Departamento de Astrofisica, Universidad de La Laguna, E-38206 La Laguna, Tenerife, Spain}

   \author[0000-0003-0975-6659]{Viggo H. Hansteen}
     \email{viggo.hansteen@astro.uio.no}
       \affiliation{Lockheed Martin Solar and Astrophysics Laboratory, Palo Alto, CA 94304, USA} 
       \affiliation{SETI Institute, 339 N Bernardo Ave Suite 200, Mountain View, CA 94043}
       \affiliation{Rosseland Centre for Solar Physics, University of Oslo,  PO Box 1029 Blindern, 0315 Oslo, Norway}
        \affiliation{Institute of Theoretical Astrophysics, University of Oslo, PO Box 1029 Blindern, 0315 Oslo, Norway}

\author[0000-0002-7788-6482]{Daniel N\'obrega-Siverio}
  \email{dnobrega@iac.es}
   \affiliation{Instituto de Astrofisica de Canarias, E-38205 La Laguna, Tenerife, Spain}
   \affiliation{Departamento de Astrofisica, Universidad de La Laguna, E-38206 La Laguna, Tenerife, Spain}
   \affiliation{Rosseland Centre for Solar Physics, University of Oslo, 
            PO Box 1029 Blindern, 0315 Oslo, Norway}
   \affiliation{Institute of Theoretical Astrophysics, University of Oslo, 
            PO Box 1029 Blindern, 0315 Oslo, Norway}

\shortauthors{}

\begin{abstract}
We present a radiation-magnetohydrodynamics (RMHD) simulation of a magnetic cancellation event. The model is calculated with the Bifrost code and spans from the uppermost convection zone to the corona. The cancellation occurs between the positive polarity of an emerged magnetic bipole and a preexisting negative polarity. We try both to understand the RMHD aspects as well as to carry out comparison to observations, in part via spectral synthesis of optically thick photospheric and chromospheric lines using the RH1.5D code, and optically thin coronal ones. The reconnection between the opposite flux systems takes place at chromospheric heights through a quasi-separatrix layer without null points. Sharp V-shaped upward-moving field lines and highly warped downward-moving post-reconnection loops are created. The chromospheric reconnection is in full swing when the colliding magnetic patches are still separated by a granular cell at the photosphere. In a later phase, photospheric cancellation takes place with submergence of the closed magnetic loops linking the opposite polarities. We carry out comparisons with the observations of the photospheric magnetic flux loss rates, as well as of the horizontal magnetic field and vertical velocity at the polarity inversion line. The reconnection outflows cause intensity brightenings, jets and different spectral features in the synthesized chromospheric spectral lines, strongly reminiscent of those found in recent observations. Coherent, twisted magnetic flux ropes are created by the flows associated with the process. Including coronal levels is crucial for proper modeling, even if no major ejection or brightening is produced in the corona in this event.
\end{abstract}


\section{Introduction}\label{sec:introduction}
Flux cancellation is one of the basic processes that determine the change of magnetic flux at the solar surface.  Widely studied from an observational point of view in the photosphere and chromosphere, there are still many open questions regarding the overall 3D structure and evolution of the cancellation sites, especially concerning the links of the low atmospheric layers to the overlying transition region (TR) and corona. In fact, a systematic study of magnetic flux cancellation processes in the framework of 3D radiation-MHD models that include photospheric, chromospheric and coronal layers has not yet been carried out; a first approach to that problem is the objective of the present paper.

In the observational literature, cancellation is often described, on the basis of photospheric or chromospheric magnetograms, as the mutual approach of opposite polarity patches until they touch each other, followed by decrease in their unsigned magnetic flux, with, sometimes, full disappearance of one of the polarities or both \citep[see, e.g., the reviews by][]{Chae_2012, Bellot_Orozco_LRSP_2019}. Many observational results along the past several decades were obtained trying to address the schematic question of whether the opposite polarities are the footpoints of an $\Omega$-loop in the process of retracting to subphotospheric levels or, rather, whether the cancellation was the result of reconnection of two initially independent flux systems, possibly leading to the detection of reconnected loops that were either submerging as $\Omega$-loops or rising as U-loops depending on the height of the reconnection site with respect to the observed level \citep{Zwaan_1987, Spruit_etal_1987}.  Yet, there is so far no unified conclusion of a configuration of rising or descending loops, nor of reconnecting opposite vertical fields, in the observed cancellation sites in different solar environments.  Early observational studies of cancellation are those of \citet{Livi_etal_1985a,Livi_etal_1989} and \citet{Martin_etal_1985}.  \citet{Harvey_etal_1999} compared the times at which several tens of cancellation events occurred at photospheric, chromospheric and coronal heights; their results strongly suggested that magnetic flux is submerging at most (perhaps all) of their cancellation sites.Subsequent high-resolution spectropolarimetric and imaging studies \citep{Chae_etal_2004, Bellot_Beck_2005, Kubo_Shimizu_2007, Fischer_2011, Kubo_etal_2010b, Kubo_etal_2014, Gosic_etal_2014, Gosic_etal_2016, Gosic_etal_2018, Kaithakkal_Solanki_2019} reported diverse and sometimes contradictory results regarding the presence of horizontal fields and plasma flows above the polarity inversion line (PIL) between the canceling polarities, which suggests the possibility that there is a variety of magnetic configurations in different cancellation sites and at different heights within each of them.  \citet{Gosic_etal_2018}, in particular, studied more than $400$ cancellation events in the interior of quiet-Sun supergranules and found, on average, short-lived (3~min) cancellation episodes in the photosphere accompanied by longer (12~min) brightenings in the chromosphere and TR. This is taken as a hint that those events are reconnection-driven, and it is suggested that the chromospheric reconnection starts before the photospheric cancellation. On the other hand, \citet{Kaithakkal_Solanki_2019} studied 11 photospheric cancellation events and found systematic differences in the LOS velocity pattern between two groups of cases: (a) both canceling polarities were established elements prior to the interaction and (b) one of the two polarities was a member of an emerging bipole. They stressed the importance of the neighboring granular motions in the cancellation process.

Photospheric magnetic flux cancellation has also been found to be related to a variety of heating, ejective or explosive phenomena on different scales, from the smallest ones such as quiet-Sun Ellerman bombs \citep{Bhatnagar_etal_2025a,Bhatnagar_etal_2025b}, or UV bursts \citep[e.g.,][and references therein]{Nelson_etal_2016,Young_etal_2018} to the ejection of surges and coronal jets \citep[e.g.][]{Chae_etal_1999, Yoshimura_etal_2003, Liu_Kurokawa_2004, Jiang_etal_2007} especially when the cancellation results from the collision of emerging flux and a preexisting field system. Cancellation has been found in relation to coronal bright points \citep{Harvey_1996, Madjarska_2019}, solar flares \citep{Livi_etal_1989, Wang_Shi_1993, Zhang_etal_2001} and blowout jets \citep[e.g.,][]{Sterling_etal_2008, Panesar_etal_2016, Panesar_etal_2017_a, Panesar_etal_2018_a, McGlasson_etal_2019, Moore_etal_2022}.

Concerning theory, a branch of the research on flux cancellation has been carried out using idealized (meaning: non-radiative) purely MHD coronal models that focus on the evolution of the magnetic field topology in a cancellation event with a view to explaining different aspects of the chromospheric and coronal heating. The papers by \citet{Priest_etal_2018, Syntelis_etal_2019} and \citet{Syntelis_Priest_2020} start from an analytical model of approaching polarities of opposite sign that lead to the formation of a current sheet in the atmosphere with magnetic reconnection across it and associated plasma heating. The relative simplicity of the model allows the authors to predict the reconnection height and the heating rate as a function of basic geometric and magnetic parameters of the problem. The model is then tested through non-radiative 2D and 3D numerical simulations. These models are an update and extension of the classical models for powering coronal bright point models through magnetic cancellation by \citet{Priest_etal_1994, Parnell_etal_1994, von_Rekowski_etal_2006a, von_Rekowski_etal_2006b, von_Rekowski_etal_2008a, von_Rekowski_etal_2008b}.  Further extensions of the basic analytical theory have been discussed by \citet{Priest_Syntelis_2021, Syntelis_Priest_2021}.  \citet{Pontin_etal_2024} have carried out an idealized purely coronal MHD 3D simulation in which the footpoints of the downward spine in a fan-spine configuration are moved at high horizontal speed toward one of the surrounding opposite-polarity patches, thus nearing a state of mutual flux cancellation. The authors study what they call the pre-cancellation state in which the approach of the opposite polarities leads to reconnection in a collapsed current sheet above the photosphere, with the emission of reconnection jets.

A second set of theoretical studies of flux cancellation have been carried out using numerical radiation-magnetohydrodynamics codes that model realistic surface convection but including only the lowermost few to several 100 km in the atmosphere.  \citet{Cheung_etal_2008} studied a process of flux emergence through the convecting layer and, in the emerged phase, identified cancellation sites at the corners between granular cells; they observed downflows with velocities of $6$ to $10$~\kms coming from a reconnection process taking place above the cancellation site, but the low height of their upper boundary ($300$~km above the photosphere) prevented them from studying the cancellation process in detail.  Other authors use an initial convectively-relaxed state and impose on it simple distributions of vertical magnetic field that change sign following checkerboard or horizontally striped patterns so as to give rise to cancellation episodes \citep{Cameron_etal_2011, Danilovic_2017, Thaler_Spruit_2017, Thaler_Borrero_2023}.  \citet{Cameron_etal_2011} found flux removal events of both the downgoing $\Omega$-loop and rising U-loop type. The flux removal occurred at a rate corresponding to an effective turbulent diffusivity of $100$-$340$~km$^2$~s$^{-1}$, with this value strongly depending on the boundary conditions. \citet{Thaler_Spruit_2017} experimented with deep boxes ($10.5$~Mm and $3.5$~Mm below the photosphere). The influence of the extended depth was negligible when measured through the equivalent turbulent diffusivity; instead, they found flux decay rates which sensitively depended on the initial strength of the vertical field. \citet{Thaler_Borrero_2023} compared simulations with initial conditions with a checkered, positive-negative vertical-field pattern to others in which flux was injected through the lower boundary and emerged through the surface. They found magnetic retraction events in the former but could not identify reconnection sites in them; on the other hand, the flux emergence cases led to reconnection events between magnetic patches resulting from the emergence at heights near the upper boundary of their numerical box ($0.64$~Mm above the surface). Finally, \citet{Danilovic_2017} discussed the close association between cancellation episodes and Ellerman bombs appearing in her numerical models, suggesting that observational constraints may sometimes prevent finding the canceling polarities at Ellerman bomb sites.  The appearance of Ellerman bombs at cancellation sites was also studied by \citet{Hansteen_etal_2017, Hansteen_etal_2019}, using Bifrost models spanning the top convection zone to the corona.  The authors discussed the relation of Ellerman bombs to UV bursts, but did not discuss the cancellation process in detail, nor study the magnetic field's evolution and topology as the opposite polarities interacted.

The extended observational and theoretical literature concerning flux cancellation reflects the fundamental importance of that phenomenon for basic aspects of solar physics like the photospheric magnetic flux budget or the heating of the chromospheric and coronal layers. The importance of flux cancellation for the former problem has been clearly shown at different levels, like globally over the solar cycle including its role in flux transport models \citep[e.g.][]{Schrijver_Harvey_1994, Mackay_Yeates_2012, Thibault_etal_2014, Cameron_Schuessler_2015}, as well as at the level of individual supergranules in the Quiet Sun \citep{Gosic_etal_2014, Gosic_etal_2016, Attie_etal_2016}.  Concerning the problem of chromospheric and coronal heating, high resolution maps from the IMaX instrument on SUNRISE indicate that the density of mixed magnetic polarity near the footpoints of hot magnetic loops may be large and that flux cancellation therefore could be an important source of energy injection into the chromosphere and corona \citep{Priest_etal_2018}.  Reconnection heating would occur much closer to the photosphere than in the case of magnetic field braiding which in numerical models occurs at small angles and throughout the coronal volume.  Flux is emerging continually over the entire solar surface. This emergence, or the random walk of already emerged flux, will result in multiple cancellations, but also in the injection of cool photospheric material into the upper chromosphere and even lower corona with observational consequences \citep{Hansteen_etal_2023} that may help in deciding on the relative importance of cancellations versus braiding in heating the outer solar atmosphere.

This paper is the first of a series in which we analyze the properties of a few cancellation events that occurred in a Bifrost numerical simulation in which magnetic flux emerged from the solar interior into an ambient magnetic field reminiscent of the Quiet Sun network, including realistic radiation transfer in the photosphere, chromosphere and corona. In this paper we focus on a cancellation event in which the positive polarity of an emerging bipole collides with a preexisting negative element, with dynamics and energetics mainly confined to the photosphere and chromosphere, leaving for a subsequent study another case in which strong TR and coronal heating leads to significant brightenings and jet-like ejections in those heights.  The layout of the paper is as follows.  Section~\ref{sec:model} outlines the methods, including a brief description of the simulation and the synthesis of optically thick photospheric and chromospheric lines.  Then, in Sect.~\ref{sec:observational_proxies}, we take a semi-observational approach, analyzing magnetograms (Sect.~\ref{sec:site_1_magnetograms}), magnetic flux evolution (Sect.~\ref{sec:flux_measurements}), field and velocity patterns at the polarity inversion line (PIL) (Sect.~\ref{sec:PIL}), and chromospheric brightenings and jet-like features from the synthetic spectra (Sect.~\ref{sec:chromospheric_brightenings_and_jets}).  Section~\ref{sec:3D_view} presents an encompassing 3D view of the cancellation site, emphasizing the reconnection patterns and phases, the formation of twisted flux ropes at the boundary between canceling domains, and the 3D distribution of variables that leads to the appearance of flames or small jets in chromospheric lines.  We finish with discussion and conclusions (Sect.~\ref{sec:discussion_and_conclusions}).

\section{Numerical setup and methods} \label{sec:model}

\subsection{The numerical model}

In order to study the cancellation phenomena we consider a 3D simulation, already used in the paper by \citet{Hansteen_etal_2019}, that includes an initial ambient field that is perturbed by significant flux emergence, and follow the evolution over a period of roughly 3 hours. The simulation was calculated using the Bifrost code \citep{Gudiksen_etal_2011}, which solves the 3D radiation-magnetohydrodynamics equations on a cartesian grid.  The initial atmosphere and magnetic field configuration come from a higher resolution version of the {\it Public Bifrost model} \citep[][]{Carlsson_etal_2016}, which contains an enhanced-network magnetic field distribution characterized by $10$-Mm long coronal loops that span the atmosphere and connect two large opposite-polarity regions at the photosphere. The modeled domain is $24\times 24\times 17$~Mm$^3$, extending $2.5$~Mm below the $<\tau_{500}>=1$ level (which is chosen as $z=0$) and $14.5$~Mm above, on a grid of $768 \times 768 \times 768$ cells giving a horizontal resolution of 31~km (as opposed to 48~km and grid of $504 \times 504 \times 496$ cells in the public model). The vertical resolution ranges from some 20~km at the bottom of the domain to less than $13$~km in the photosphere, chromosphere, transition region, and lower corona, with the grid spacing increasing between $13$~km and $80$~km in the range $z=(3.5,14.5)$~Mm.

At the start of the simulation a horizontal flux sheet with a strength of $B_y=2000$~G spanning the region $x=(3,16)$~Mm is injected in the sense that the horizontal magnetic field is imposed in upflow regions at the bottom boundary and transported by already existing flows, as previously done in other simulations \citep[e.g.,][]{Stein_Nordlund_2012,Archontis_Hansteen_2014}. This injection is maintained for a period of 100~minutes before being ramped down.  As the injected field rises through the convection zone and emerges through the photosphere, several magnetic bipoles appear and are subsequently transported by granular convective motions, eventually interacting (and indeed sometimes canceling) with the preexisting ambient field and/or other newly emerged magnetic elements.  We have identified five sites of vigorous interaction where cancellation occurs as opposite photospheric polarities are driven together by these flows. Analysis of one of them forms the main thrust of this study, while a second case, with different reconnection patterns and coronal response, will be studied in a forthcoming~paper.

\subsection{Spectral synthesis}\label{sec:spectral_synthesis}

For direct comparison with the observations, we provide in this paper spectral synthesis of optically thick lines computed with the RH1.5D code \citep{Pereira_Uitenbroek2015}. This code solves the equation of radiative transfer for multidimensional models of stellar atmospheres, considering non-local thermodynamic equilibrium (NLTE) and partial redistribution of the scattered radiation. We synthesized the \CaIIIR, \CaIIK, \MgIIk, and \MgIItriplet\ lines, so as to get good diagnostics at different chromospheric heights \citep[see][]{Leenaarts_etal_2013,Pereira_etal_2015,Bjorgen_etal_2018}.  For the photosphere, we calculated the four Stokes profiles for the \FeIHMIline\ line; an observational value for the field strength is then calculated from the Stokes-V spectra using standard procedures \citep[e.g.][]{2003ApJ...592.1225U}, which is then used to produce synthetic magnetograms. We also calculated synthetic FUV and EUV lines (like \SiIV, \FeIX, \FeXII\ and \FeXV) using the CHIANTI package \citep{Dere_etal_2023}, to monitor any TR or coronal response associated with the cancellation.

\subsection{The patch and PIL selection algorithm}\label{sec:DBSCAN}

For the selection of relevant magnetic patches and PIL neighborhoods in Sections~\ref{sec:flux_measurements} and \ref{sec:PIL} we use the Density-Based Spatial Clustering of Applications with Noise algorithm \citep[DBSCAN;][]{Ester_etal_1996, Schubert_etal_2017}. This is an unsupervised machine-learning technique that identifies clusters as dense regions in the data space.  It is implemented in Scikit-learn \citep{Pedregosa_etal_2011}. DBSCAN depends on two main parameters: $\epsilon$, the maximum distance between two points for them to be considered neighbors; and $\min_{\mathrm{samples}}$, the minimum number of points required within the $\epsilon$ radius to form a dense region.  DBSCAN handles arbitrarily shaped clusters effectively, and is thus well suited to identify the magnetic patches or PIL regions involved in the cancellation events.

\section{Results: Observational proxies}
\label{sec:observational_proxies} 

The cancellation site chosen for in-depth study in this paper leads to cancellation phenomena mostly in the lower atmospheric layers. In this section, we provide different observational proxies that allow us to characterize this site as a standard case among those found in the observational literature.

\begin{figure*}[htbp]
\centerline{\includegraphics[width=\textwidth]{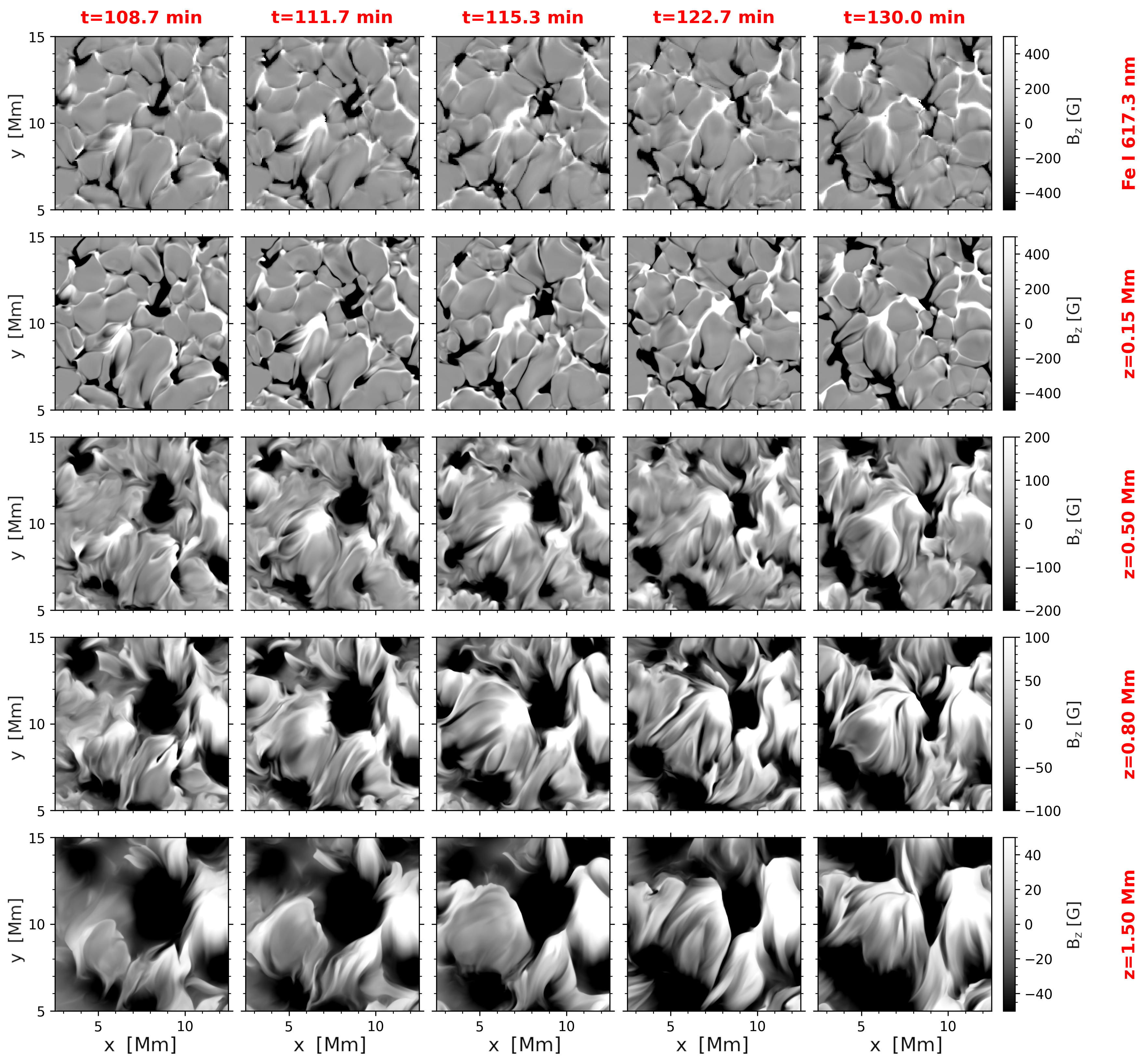}}
\caption{Magnetograms taken at different instants along the cancellation
  process. Top row: synthetic magnetograms obtained through the
  V-Stokes profile of the \FeIHMIline\ line. Lower rows: horizontal maps of
  the vertical magnetic field component at heights $z=0.15$~Mm (second row),
  $0.5$~Mm (third row), $0.8$~Mm (fourth row), and $1.5$~Mm (bottom row).
  The accompanying animation 
shows the time evolution for rows 2 and 4 of this figure between $t=101.7$~min and $t=130.0$~min.
  (An associated animation is available for this figure.)
  \label{fig:cancellation_site_1_general}}
\end{figure*}

\subsection{Event description on the basis of magnetograms }\label{sec:site_1_magnetograms}

In this section we consider the evolution at individual heights of the opposite polarities that advance toward each other and end up showing a clear cancellation behavior.  Figure~\ref{fig:cancellation_site_1_general} shows (first row) the synthetic magnetograms calculated from the \FeIHMIline\ line (Sect.~\ref{sec:spectral_synthesis}), followed by maps of the vertical component of the magnetic field on horizontal cuts in the simulation box at heights $0.15$, $0.50$, $0.80$, and $1.5$~Mm above $z=0$ (rows 2 through 5). For speed, we will also refer to these horizontal $B_z$-maps as magnetograms.  In fact, the synthetic magnetogram (top row) and the horizontal $B_z$-map at $z=0.15$~Mm are very similar, as expected given that the wings of the \FeIHMIline\ line are formed roughly in the range $100$-$150$~km \citep[see][]{Norton_etal_2006}.

The cancellation in this site directly occurs in the aftermath of a process of flux emergence. In the low photospheric levels (uppermost two rows), the magnetic field distribution is seen to approximately follow the granulation pattern, with strong field concentrations interspersed in it. In the initial time shown (left column $t=108.7$~min), the bipole resulting from the emergence is seen in the bottom left corner of the panel, centered at about $(x,y)=(5,8)$~Mm (see also the accompanying animation). Its opposite polarities separate from each other roughly following a diagonal in the frame of the figure, with the positive polarity traveling toward a preexisting, stable negative polarity located at around $(x,y)=(9,12)$~Mm.  By $t=111.7$~min (second column), the two approaching opposite polarities are facing each other and a {\it strait} of weakly magnetized plasma is seen to separate them, which is, in fact, occupied by a single granule. The mutual approach continues and by $t=115.3$~min (third column) the two strong-field patches are achieving contact. A direct interpretation of these facts, disregarding the events in the upper layers, would state that the cancellation process starts at this point. The cancellation continues in the subsequent $\approx 15$~min (fourth and fifth column), with a clear reduction of the apparent size of the polarities, even though they do not disappear completely in the process; this is partly due to the fact that they merge with other polarities that approach the cancellation site from different directions.

At higher levels ($z=0.5$, $0.8$~and~$1.5$~Mm, third to fifth rows in the figure), the two approaching polarities are in contact the earlier the higher in the photosphere/chromosphere: in fact, upwards of $z=0.8$~Mm the contact is established before the earliest time shown in the figure ($t=108.3$~min). This is in part a consequence of the fanning-out shape of the established negative polarity, which is seen to occupy a larger area than in the lower level. The apparent cancellation proceeds in this case during the whole time evolution shown in the figure, and the interface between the two polarities is sharper in these levels than in the lower level. All of the foregoing is a first indication that the events leading to and accompanying the cancellation may have different geometries, connectivity and appearance at the different heights, so that an encompassing 3D study is necessary.

\subsection{Flux measurements}\label{sec:flux_measurements}

The determination of the time evolution of the magnetic flux in the canceling patches is fraught with difficulties, both in observational and numerical simulation contexts, especially when, like in the present case, the canceling patches split or merge with neighboring polarities. Different methods have been used in the literature of the past decades; particularly refined is the technique presented by \citet{Gosic_etal_2014, Gosic_etal_2018}, who used the YAFTA method \citep{Welsch_Longcope_Yafta_2003} including additional improvements to maximize the accuracy in the determination of the canceling flux.

\begin{figure}[htbp]
\centering
\centerline{\hskip 0mm 
\includegraphics[width=0.48\textwidth]{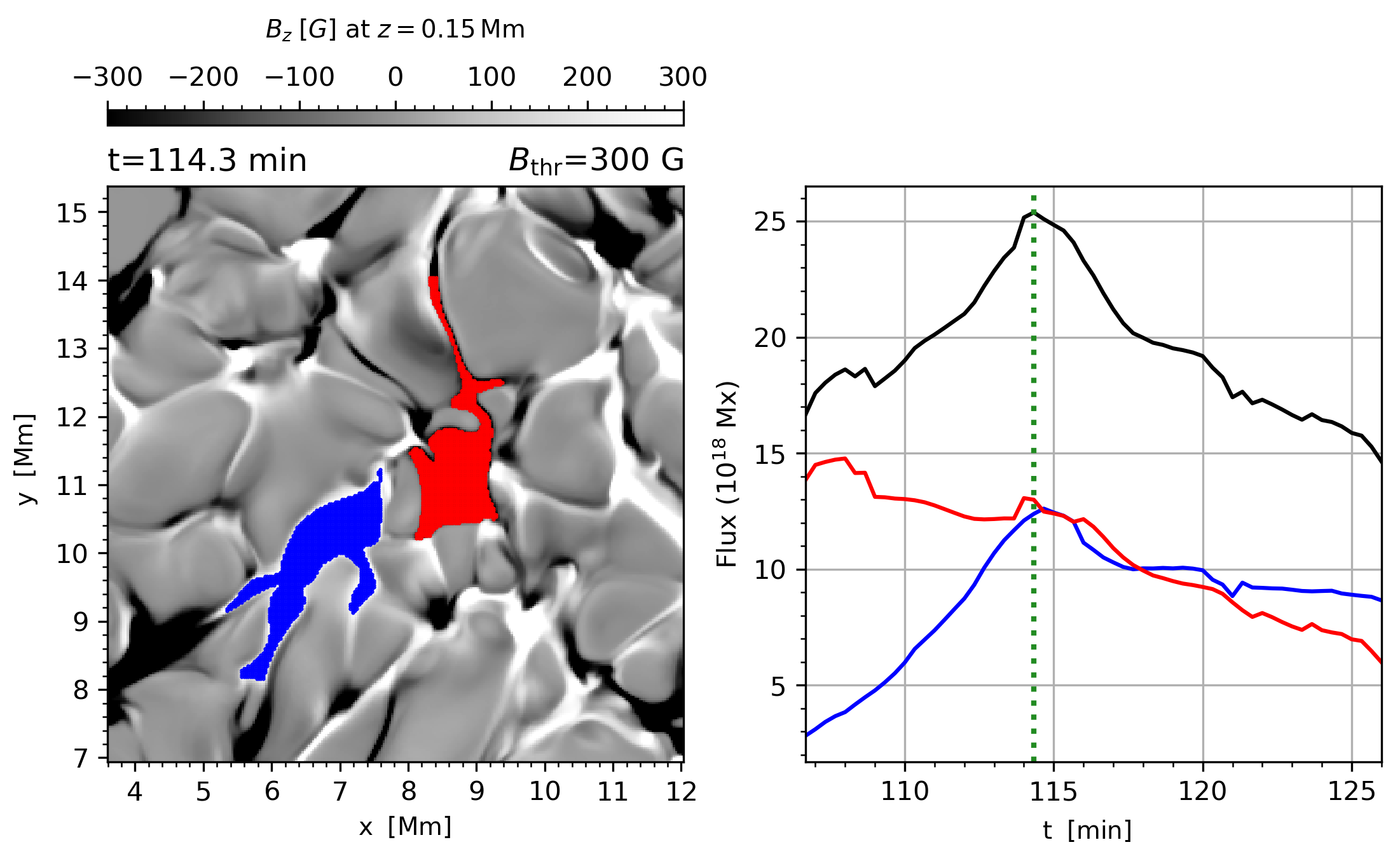}}
\caption{Canceling flux patch detection on a horizontal cut at $z=0.15$~Mm
with $B_{\mathrm{thr}}=300$~G. Left: Vertical field strength map
  (grey scale) with, superimposed, the identified canceling patches
  (red: negative polarity, blue: positive polarity). Right:
  Time evolution of the integrated unsigned magnetic flux in the positive
  patch (blue), negative patch (red) and their sum (black).
  The time evolution of the flux patches between $t=106.7$~min to $t=126.0$~min
  can be seen in the accompanying animation.
  (An associated animation is available for this figure.)
\label{fig:flux_patches}}
\end{figure}

In the present simulation, we find strong canceling polarities that are not well-isolated, roundish patches throughout their lifetime.  To accurately characterize their flux evolution, we employ the following method: we first identify pixels at $z = 0.15$~Mm with magnetic field strength above a threshold, namely $B_{\mathrm{thr}} = 300$~G; we then apply the DBSCAN algorithm (Sect.~\ref{sec:DBSCAN}) with parameters $\epsilon = 0.063$ Mm and $\min_{\mathrm{samples}} = 5$, which effectively identifies the pixels associated with the two main opposite polarities undergoing cancellation.  The results are shown in Fig.~\ref{fig:flux_patches} and the accompanying animation.

The left panel shows the identified patches for the positive (blue) and negative (red) polarities; the right panel contains the time evolution of the flux within the identified patches along the cancellation process.  The identification is unambiguous in most snapshots, although occasionally a nearby extension exceeds the field strength threshold and is temporarily included in the patch, causing minor fluctuations in the flux curve that do not affect our main results.  We see that, as expected from the results of the foregoing section, the positive polarity, which is created by an emerging process, increases its flux until $t=115$~min; from then on, both polarities decrease their unsigned flux for at least $10$~min. The average unsigned flux decay rate per polarity is $\pot{2.7}{19}$~Mx~h$^{-1}$.  Both this flux decay rate and the magnetic flux values of Fig.~\ref{fig:flux_patches} fall well within the range obtained by \citet{Park_etal_2009} following observations with the Hinode/SOT instrument of canceling magnetic features.

\begin{figure}[htbp]
\hbox to \hsize{\hskip -1.2cm \includegraphics[width=0.61\textwidth]{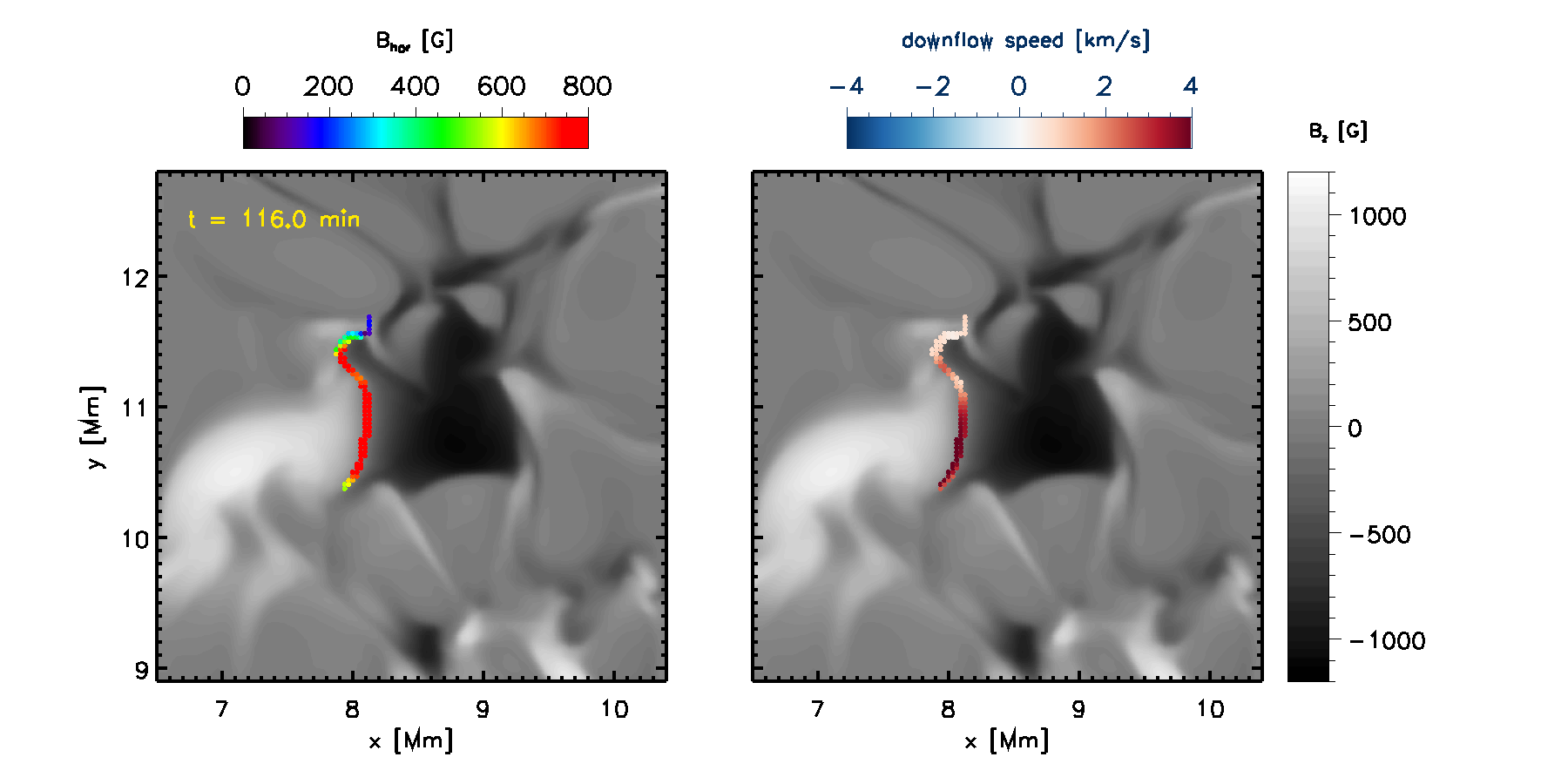}\hfill\hskip 1cm\ }
\vskip -3mm
\hbox to \hsize{\hskip -6mm
\vbox{\hsize 0.25\textwidth 
\includegraphics[width=0.25\textwidth]{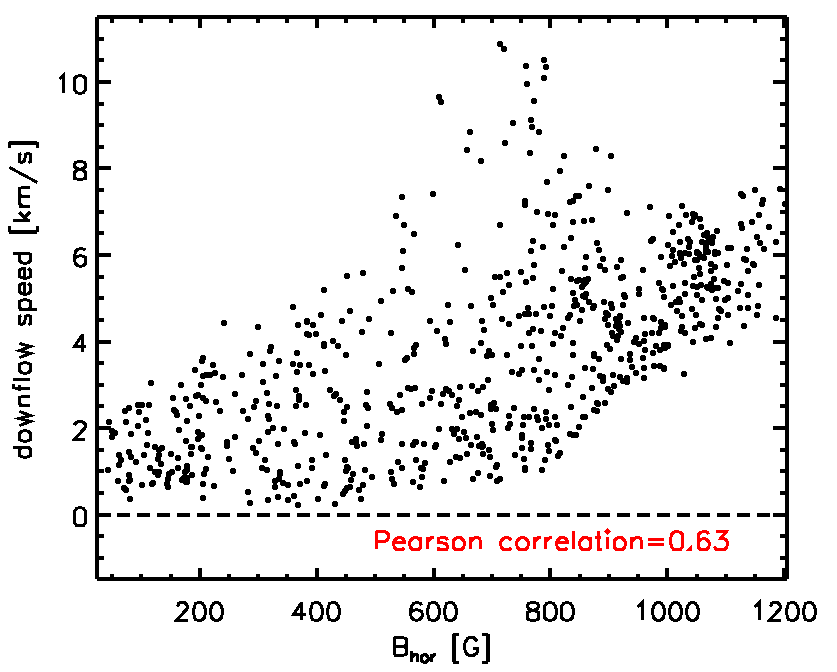}}
\hskip -2mm\vbox{\hsize 0.25\textwidth 
\includegraphics[width=0.24\textwidth]{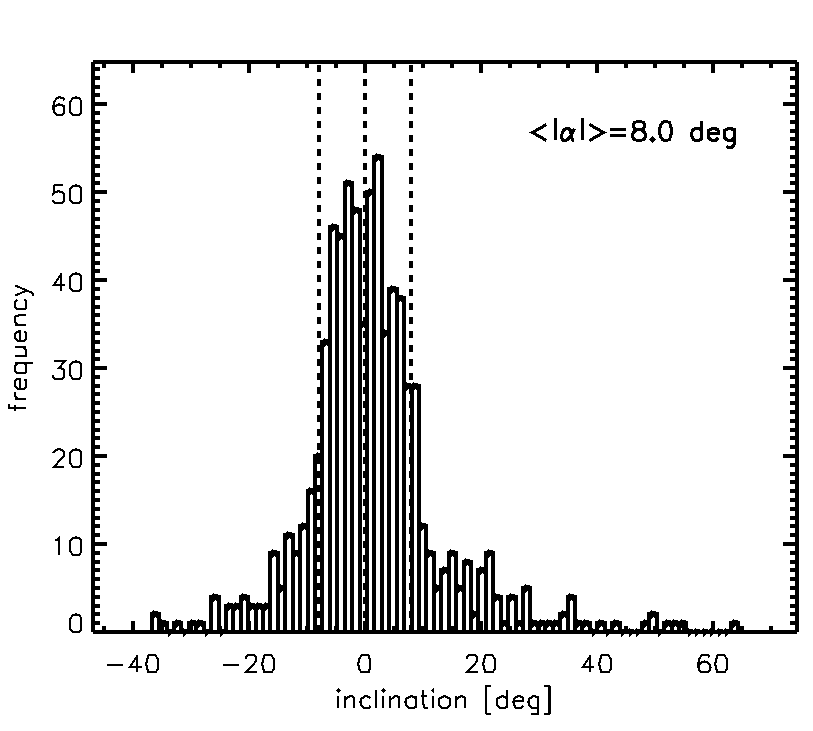}}
\hfill}
\hbox to \hsize{\hskip -1.2cm\includegraphics[width=0.61\textwidth]{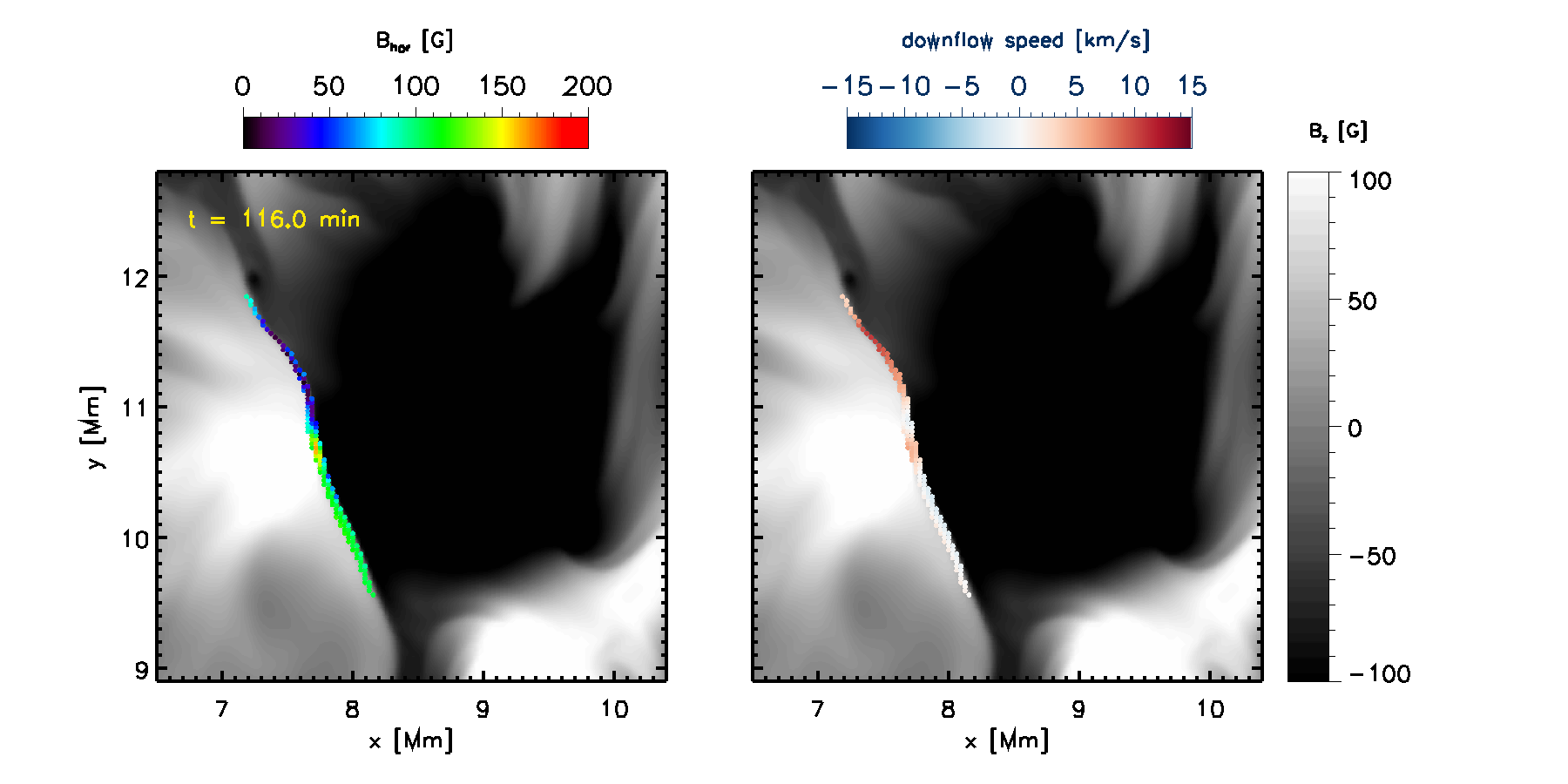}\hfill}
\caption{Top row: the PIL between the canceling patches on a horizontal cut at $z=0.15$~Mm at $t=116.0$~min. The pixels on the PIL are colored based on the horizontal field strength $\Bhor$ (left) and downflow (right), with colorscale given at the top of the panels.  The background map shows the vertical magnetic field, $B_z$ (colorbar on the right). The associated animation shows the evolution of those two panels in the time range $t=116.0 - 122.0$~min including average values for $\Bhor$, downflow speed, and the inclination angle to the horizontal $\alpha$ for the PIL in each snapshot. Middle row: cumulative diagrams for the pixels in the PIL shown in the animation: left, scatter plot of downflow vs $\Bhor$; right: histogram for the inclination angles; the range $\pm\langle|\alpha|\rangle$ is indicated with dashed vertical lines. Lower row: same as the top row, but now for $z=1.0$~Mm.  (An associated animation is available for this figure.)
  \label{fig:PIL_z=0.15}}
\end{figure}

\subsection{Magnetic field strength and downflows in the Polarity Inversion
  Line (PIL)} \label{sec:PIL}

In the observational literature on magnetic flux cancellation, particular attention has been devoted to measurements along the line of sight on top of the polarity inversion line (PIL) between the canceling polarities \citep{Bellot_Beck_2005, Chae_etal_2004, Chae_etal_2010, Gosic_etal_2018, Kaithakkal_Solanki_2019}.  In this section we try to approach this kind of measurements based on the simulation data.

To detect PILs in the simulation, we first take a magnetogram at a given height and apply a 2$\times$2 spatial binning.  Then, we look for locations where pixels with sufficiently strong positive and negative field strengths coexist in the immediate neighborhood.  In detail, we use squares of 5-pixel side as neighborhoods around every single pixel and set a positive threshold $B_{\mathrm{PIL}}$. When in a given square at least three pixels have $B_z >B_{\mathrm{PIL}}$ and at least three other pixels have $B_z < -B_{\mathrm{PIL}}$, then the central pixel of the square is declared to belong to the PIL.  This algorithm has been seen to correctly detect the interface between opposite polarities, not just in the main cancellation site but in other secondary opposite-polarity pairs.  To select the PIL of interest, first we restrict the field of view to the region containing the canceling polarities and then the DBSCAN procedure, with $\epsilon=0.3$~Mm and $\min_{\mathrm{samples}}=1$ (Sect.~\ref{sec:DBSCAN}), is applied. To ensure continuity of the selected PIL, the grouping calculated by the DBSCAN is determined by using together, for each snapshot, the previous and subsequent ones, so that merging or splitting fragments are consistently grouped.  A few segments not belonging to the PIL of the cancellation site that the DBSCAN was not able to separate were manually removed.

Figure~\ref{fig:PIL_z=0.15} (top row) and the associated animation show the PIL determined using magnetic data at $z=0.15$~Mm with $B_{\mathrm{PIL}} = 100$~G. The instant shown in the figure, $t=116.0$~min, corresponds to the beginning of the main photospheric cancellation phase shown in Fig.~\ref{fig:cancellation_site_1_general}; the accompanying animation shows the evolution between $t=116.0$~min and $t=122.0$~min with cadence $10$~s. The colored dots show the values of horizontal field strength, $\Bhor$ (left panel), and downflow speed (right panel), on the identified points of the PIL. We see that the PIL at photospheric levels is characterized by high values of $\Bhor$ [up to O$(1$~kG$)$] and predominant downflows of a few to several~\kms, not just in the figure but along the whole evolution (as apparent in the accompanying animation). The average value of $\Bhor$ and downflow speed at all pixels in the PIL for the snapshots in the animation are $630$~G and $3.7$~\kms, respectively.  These averages are fully within the observed range for $\Bhor$ and not far from it for the downflow speed, as will be discussed in Sect.~\ref{sec:discussion_and_conclusions}.4.  The predominance of strong horizontal field and downflows is corroborated in the middle row: the left panel contains a scatter plot for $\Bhor$ and downflow speed in the PIL in the snapshots in the animation; there is in fact a moderately good correlation (Pearson correlation coefficient $0.63$) between those quantities.  The right panel contains a histogram for the inclination angle $\alpha = \hbox{atan}(B_z/\Bhor)$ showing that the magnetic field is not far from horizontal in the majority of pixels in the PIL.

The lower row of Fig.~\ref{fig:PIL_z=0.15} shows the corresponding panels for cuts at a mid-chromospheric height ($z\hskip -3pt =\hskip -3pt1.0$~Mm). We still see a sharp PIL and predominance of downflows. A scatter plot (not shown) for $\Bhor$ and downflow speed at \hbox{$z\hskip -3pt =\hskip -3pt 1.0$~Mm} for the same time series as in the animation reveals a weaker correlation (Pearson $0.43$). The average of $\Bhor$ and downflow speed for the same snapshots as in the animation is now $83$~G and $5.6$~\kms, respectively.

\subsection{Chromospheric brightenings and jets associated with the cancellation process} \label{sec:chromospheric_brightenings_and_jets}

\begin{figure*}[htbp]
\centering
\vbox{
\centerline{\hskip -4mm \includegraphics[width=1.2\textwidth]{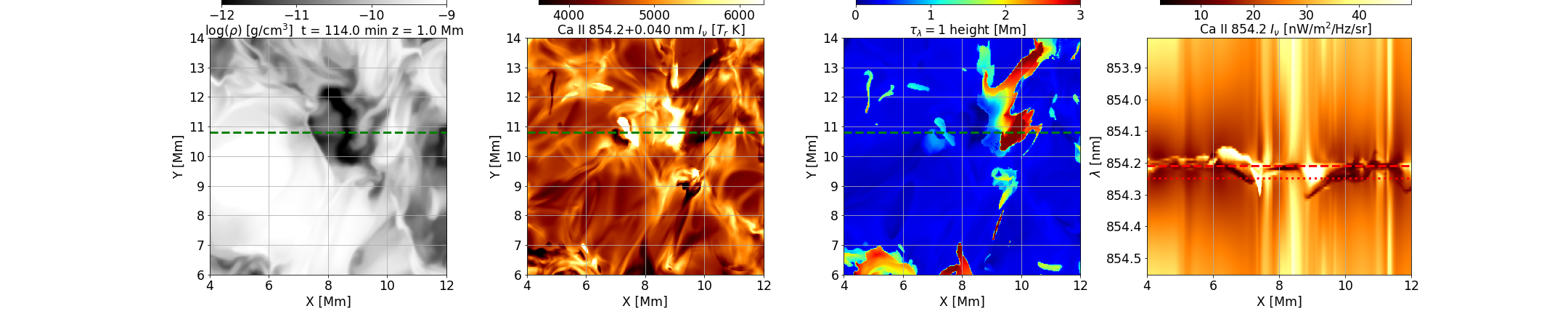}\hfill}
\centerline{\hskip -4mm \includegraphics[width=1.2\textwidth]{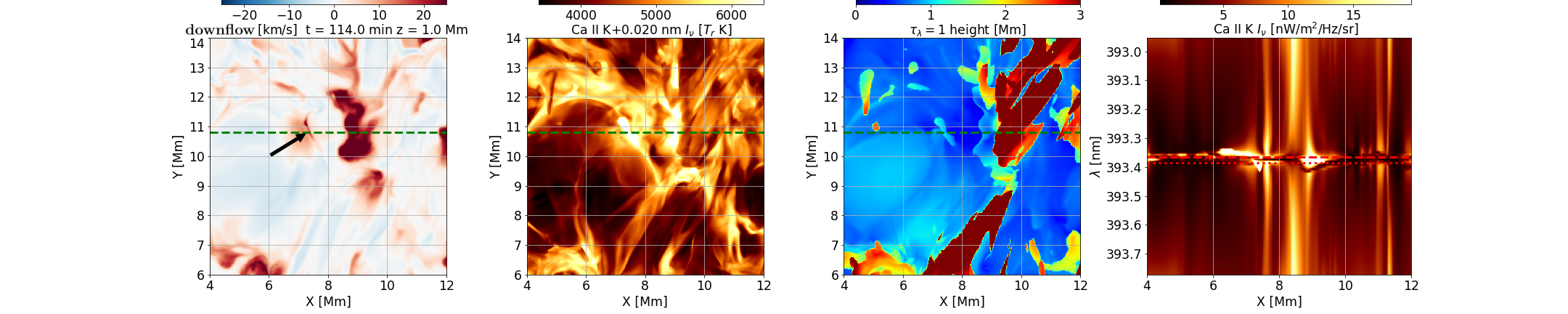}\hfill}
\centerline{\hskip -4mm \includegraphics[width=1.2\textwidth]{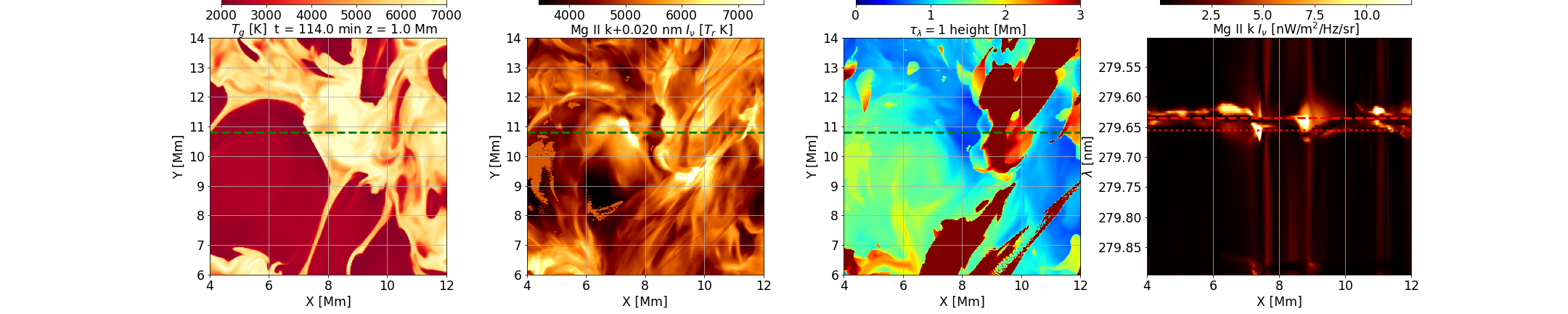}\hfill}
\centerline{\hskip -4mm \includegraphics[width=1.2\textwidth]{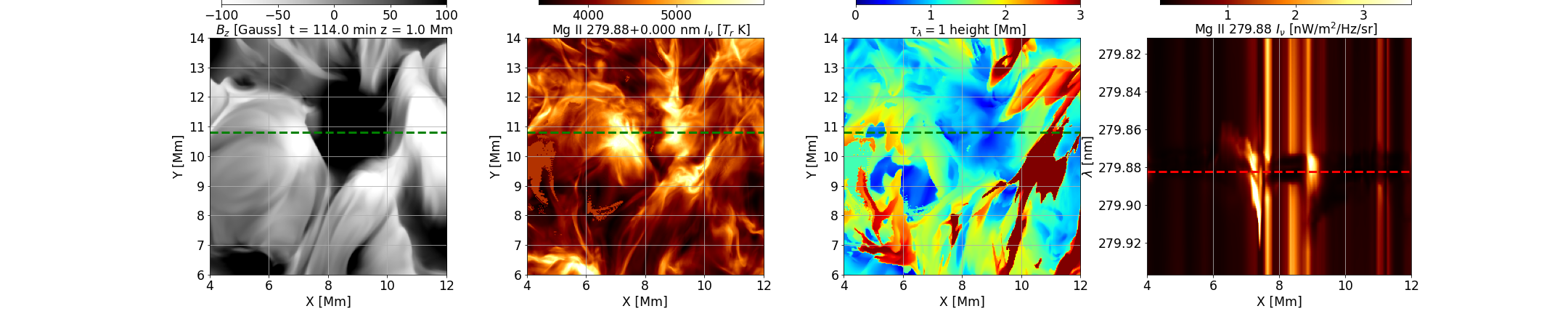}\hfill}
}
\caption{Result of the synthesis of the \CaIIIR, \CaIIK, \MgIIk, \MgIItriplet\ lines in the vicinity of the cancellation site. The top row shows, from left to right: the density at 1~Mm; filtergrams of the \CaIIIR~line at $0.04$~nm from line center near the $k_{2r}$ peak; the formation height at this wavelength; and the line spectra along the cut $y=10.8$~Mm (shown by the green dashed line). The red dashed line is placed at line center, while the red dotted line shows the wavelength plotted in the second column. The second and third rows show the same quantities for the \CaIIK~and \MgIIk~lines, but at $0.02$~nm from line center, with the downflow speed and temperature at 1~Mm replacing the density in the left column. Finally, the fourth row shows these quantities for the \MgIItriplet~line, with the vertical magnetic field $B_z$ at a height of 1.0~Mm being shown in the left column. The black arrow in the vertical velocity figure shows the location of the ``ridge'' discussed in Sect.~\ref{sec:chromospheric_brightenings_and_jets}.
\label{fig:spec_site1_ca8542_caK_mgk}}
\end{figure*}

What are the observational signatures of the cancellation in the chromosphere?  To address this question, we use Fig.~\ref{fig:spec_site1_ca8542_caK_mgk}, which shows various quantities in the area around the cancellation site at $t=114.0$~min, i.e., at a stage just before column~3 of Fig.~\ref{fig:cancellation_site_1_general} in which the opposite polarities are about to touch at the photosphere.  The emerging bipole discussed in Sect.~\ref{sec:site_1_magnetograms} has brought dense magnetized material up into the chromosphere in a bubble-shaped structure centered at $(x,y)=(5,8)$~Mm (see, e.g., the density $\rho$ and $B_z$ panels in the first column of Fig.~\ref{fig:spec_site1_ca8542_caK_mgk}).  The density in the preexisting flux region is much lower than in the newly emerged flux, which is carrying significant mass into the chromosphere. In contrast to the density we find that the emerging plasma is much cooler, of order $3000$~K or less, than what is found in the pre-chromosphere at $z=1$~Mm, close to $7000$~K. As these fields converge, a sharp jump in density forms, demarcating the two approaching magnetic flux systems. It is along this jump that reconnection proceeds as the systems are pushed together. There, we find material that is flowing rapidly downward, at $\simeq 20$~\kms\ (see the downflow panel in the 2nd row, first column) and forms a ridge-shaped structure at the location of the jump (marked by a black arrow in the panel)

The second column of Fig.~\ref{fig:spec_site1_ca8542_caK_mgk} contains the intensity integrated along the vertical line-of-sight of \CaIIIR\ at $0.04$~nm (or $14$~\kms Doppler shift) from line center, \CaIIK\ at $0.02$~nm (or $15$~\kms Doppler shift) from line center, \MgIIk\ at $0.02$~nm (or $21.4$~\kms Doppler shift) from line center, and \MgIItriplet\ at line center; while the third column shows the corrugated formation height of those lines at the given wavelengths, or, more precisely, the height at which $\tau_\lambda=1$ is reached for each of them. The $\tau_\lambda=1$ layers for these particular wavelengths of \CaIIIR, \CaIIK, and \MgIIk\ at the ridge location are all located at roughly $\sim 1$~Mm above the nominal height of the photosphere.  The ridge is visible in all three lines as an elongated intensity enhancement at these wavelengths as the high-velocity, high-chromospheric material at $1$~Mm generates higher intensities than in the regions surrounding the ridge, which are all formed at lower heights near the temperature minimum with lower velocities. We note that the high density, but cold and draining, newly emerged plasma situated to the left of the ridge also has high opacity and a formation height close to 1~Mm, but does not become as bright as the plasma forming the ridge. We do not see any significant chromospheric temperature rise in the ridge at this height. As the cancellation proceeds, the density jump moves to the right in the positive $x$-direction along with the current sheet between the magnetic systems which remains bright in the red wings of \CaIIIR, \CaIIK\ and \MgIIk.

Further information on the cancellation process can be gathered from the line profiles of the calcium and magnesium lines obtained along the cut $y=10.8$~Mm shown in the fourth column of Fig.~\ref{fig:spec_site1_ca8542_caK_mgk}. Those panels show a red ``spike'' at the location of the PIL for all four lines, indicating downflowing velocities of order 20~\kms which correspond to the velocity of the ridge at $1$~Mm apparent in the left panel of the second row.  To the left of this spike in the spectrum, at lower values of $x$, extended blue-shifted emission is found in the \CaIIIR, \CaIIK, and \MgIIk\ lines (but not the \MgIItriplet\ line) of similar though slightly lower velocity amplitude than in the red spike. At the time of this figure the blue emission is formed at greater height than 1~Mm and is therefore not visible in the row 2 velocity map. The scenario here is similar to that reported in observations. For instance, \citet{Gosic_etal_2018} find that the observed temporal evolution of photospheric cancellation regions implies that chromospheric reconnection is a plausible scenario: asymmetries in the line profiles show evidence of both up- and downflowing plasma while local brightenings in chromospheric lines result, in their interpretation, from the energy release. Indeed, in the line profiles computed here we can see both sides of what appears to be a bi-directional jet in these lines, and their relationship with the reconnection site will become clear in the sections that follow which describe the 3D structure of this cancellation event. In addition to the bright red spike and blue excursions we also find symmetrical ``moustache''-like spectra, with higher than average emission at all wavelengths in all but the line core in the four lines in the close vicinity ($\sim 100$~km) of the cancellation site.

\begin{figure}[htbp]
\centering
\centerline{\includegraphics[width=0.50\textwidth]{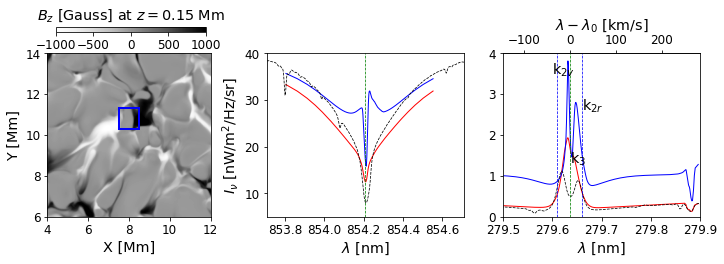}
}
\caption{Chromospheric line profiles during the cancellation event.  Vertical ($B_z$) magnetic field at $z = 150$ km (left panel), average profiles of the \CaIIIR\ (middle panel) and \MgIIk\ (right panel) lines for the entire computational domain (red curve) and in the small patch centered on the cancellation site (blue curve) indicated by the blue rectangle in the left panel.  The line centers of \CaIIIR\ and \MgIIk\ are marked with green vertical dashed lines.
\label{fig:cancellation_site_1_line_profiles}}
\end{figure}

Line plots of the line profiles of the chromospheric \CaIIIR\ and \MgIIk\ lines from central parts of the cancellation site can provide further insight and are shown in Fig.~\ref{fig:cancellation_site_1_line_profiles} next to the photospheric ``observable'' $B_z$.  For both lines, the profile averages for the entire computational domain are shown in red, while the average profiles in a $1\times1$~Mm$^2$ box centered on the cancellation site are shown in blue (the box is marked with a blue solid line in the leftmost panel). For further comparison with observations of the \CaIIIR\ line we have plotted (dashed line, middle panel) the average disk-center \CaIIIR\ profile observed by \citet{Brault_Neckel_1987}.  In the right panel we plot a typical average Quiet Sun \MgIIk\ line profile measured with the Interface Region Imaging Spectrograph (IRIS) \citep[][]{2014SoPh..289.2733D} taken on Feb 25, 2014 at 18:59:47~UT (dashed line, right panel). Finally, the right panel also shows the locations of the k$_{2v}$, k$_{2r}$ and k$_3$ peaks. The equivalent peaks in the \CaIIK\ line are named with a capital K, and we will use the same nomenclature for the peaks found in the \CaIIIR\ line. As the observed profiles presented in the figure represent quiet Sun regions, we do not necessarily expect them to match the profiles of the emerging flux region modeled here; they are included mainly for context. However, we note that the width of the core of the \CaIIIR\ line is set by the turbulent velocities found in the lower to mid chromosphere, and thus sensitive to the spatial resolution of the model with a broader core arising at higher spatial resolution. A grid cell resolution of order 30~km or better seems sufficient to reproduce observed widths. On the other hand, the width of the \MgIIk\ core is set by either the turbulent velocity of the upper chromosphere, or alternately, by the amount of mass found in the chromosphere. In the present case, increasing the grid cell resolution does not impact the core width greatly, flux emergence leads to an increase of the amount of material in the upper chromosphere and thus line widths not too different from those observed. These points are discussed in greater detail by \citet{Hansteen_etal_2023,Ondratscheck_etal_2024} as well as in Sect.~\ref{sec:discussion_and_conclusions} of this paper.

For both line spectra, we note that the profiles are on average brighter at all wavelengths in the canceling region than what is found in the profile average over the computational box. The Mg~II~k$_2$ peaks as well as the equivalent locations in the \CaIIIR\ line spectrum are bright and show large asymmetries. In particular, we find that the red K$_{2r}$ peak of the \CaIIIR\ line is brighter than K$_{2v}$. Referring back to Fig.~\ref{fig:spec_site1_ca8542_caK_mgk} we identify this peak with the ``spike'' discussed in connection with that figure. We have considered the time evolution of the \CaIIIR\ spectrum and find that, averaged over a $1\times 1$~Mm$^2$ sized region centered on the cancellation site, this red asymmetry lasts throughout the canceling episode. Figure~\ref{fig:cancellation_site_1_line_profiles} shows that the shape of the \CaIIIR~intensity profile in the central cancellation region is very similar to that observed by \citet{Gosic_etal_2018}, even though cancellation in the current paper occurs in a stronger field environment than in the quiet Sun.  This can be ascertained by comparing our Fig.~\ref{fig:cancellation_site_1_line_profiles} (or the 4th column of Fig.~\ref{fig:spec_site1_ca8542_caK_mgk}) with their Figures~9 and~10, especially with regard to the asymmetric emission in the red $K_{2r}$ peak of \CaIIIR, as well as the enhanced brightness at all wavelengths.  On the other hand, the shape of the spatially averaged \MgIIk\ profile and its asymmetry appear much more sporadic in time and we find short-lived intensity enhancements in which the blue peak dominates the spatially averaged spectrum, as can be seen in the right panel. Returning to Fig.~\ref{fig:spec_site1_ca8542_caK_mgk} it is clear that these blue enhancements preferentially occur in a region (well) to the left of the red spike in all three chromospheric lines.

\begin{figure}[b]
\centerline{\includegraphics[width=0.5\textwidth]{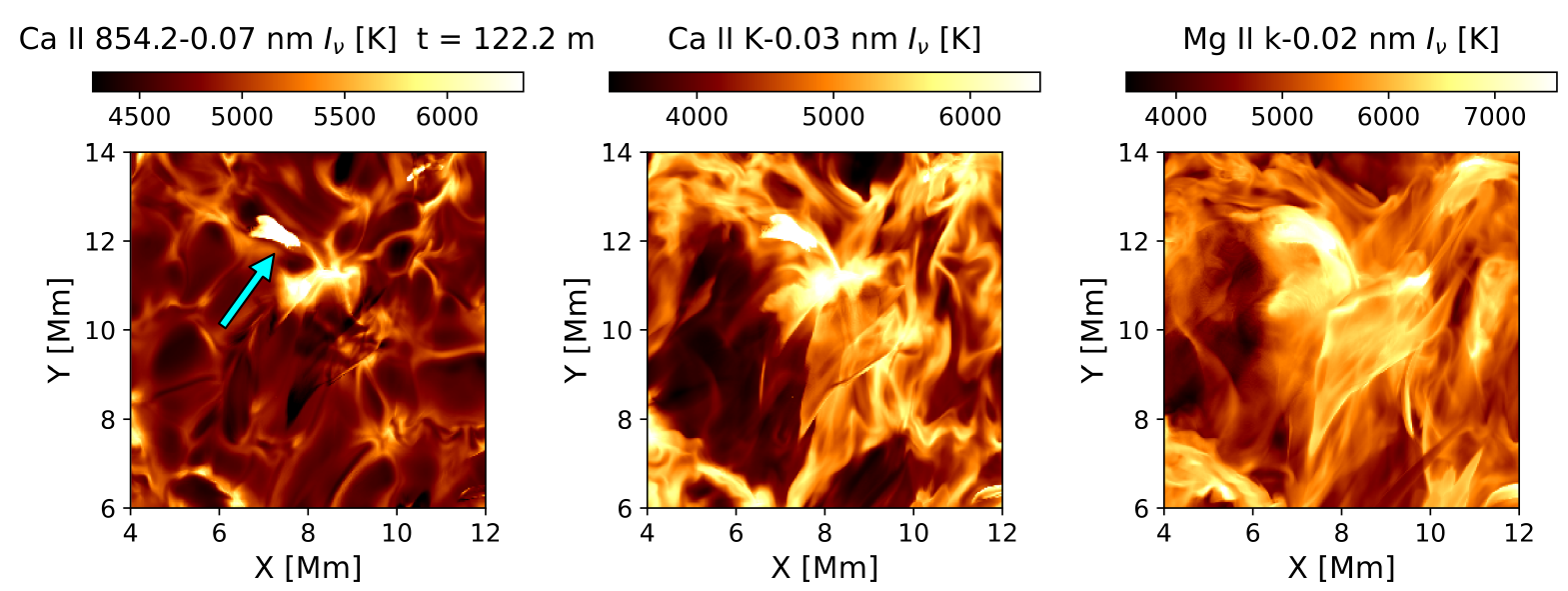}}
\caption{ Emission from a reconnection-driven jet (cyan arrow in the left panel) above the cancellation site in the blue wings of the \CaIIIR, \CaIIK, and \MgIIk~lines. The images show filtergrams at a wavelength 25~km s$^{-1}$ from the line center in each of these lines.  The time evolution of this region, including the jet, in the \CaIIIR~line can be seen in the accompanying animation.  (An associated animation is available for this figure.)
\label{fig:ca8542_caK_mgk3_flame}}
\end{figure}

While the red side of the core profile has very distinctive features we find that there are also clear signals of the cancellation event in the blue wings of the chromospheric diagnostics in this model: Considering emission in the blue wings of the chromospheric lines, to the left of the cancellation site we find several `jets' extending away from it. These jets appear sporadically and last for some tens of seconds, up to a few minutes, throughout the cancellation event. For example (Fig.~\ref{fig:ca8542_caK_mgk3_flame}), halfway through the event, some $700$~sec after the first enhanced emission is visible, we find a small jet originating in the PIL that is visible in all three chromospheric lines \CaIIIR, \CaIIK, and \MgIIk. The figure shows a snapshot of this jet some $40$~sec after it first becomes visible in these chromospheric lines at a wavelength corresponding to 25~\kms in their blue wings. This particular jet lasts $120$~sec before it becomes invisible, moving rapidly to the left in the figure.  The formation height ($\tau_\lambda = 1$) of the jet emission is $\sim 1.8$~Mm above the photosphere for both the calcium lines and slightly higher $\sim 2$~Mm for magnesium. The regions in the immediate vicinity of the jet at this $\lambda$ are formed much lower; $< 500$~km for \CaIIIR, approximately 600~km for \CaIIK, and 1~Mm for \MgIIk. Thus, the cancellation site is bounded by two high velocity regions: The red high velocity spike that occurs just above the PIL and the blue jet(s) that form sporadically to the left, moving at high velocity both horizontally and vertically and pointing away from the reconnection region.

As should be evident from the above, the interactions occurring in the
cancellation site are highly dynamic and clearly detectable in the line profiles
of the chromospheric calcium and magnesium lines. However, this
  particular cancellation site does not have any counterpart in lines
formed at TR or coronal temperatures.

\begin{figure*}[htbp]

\centerline{\ \hskip -0.5mm
\vbox{\hsize=0.45\textwidth
\hskip 1mm\includegraphics[width=0.445\textwidth]{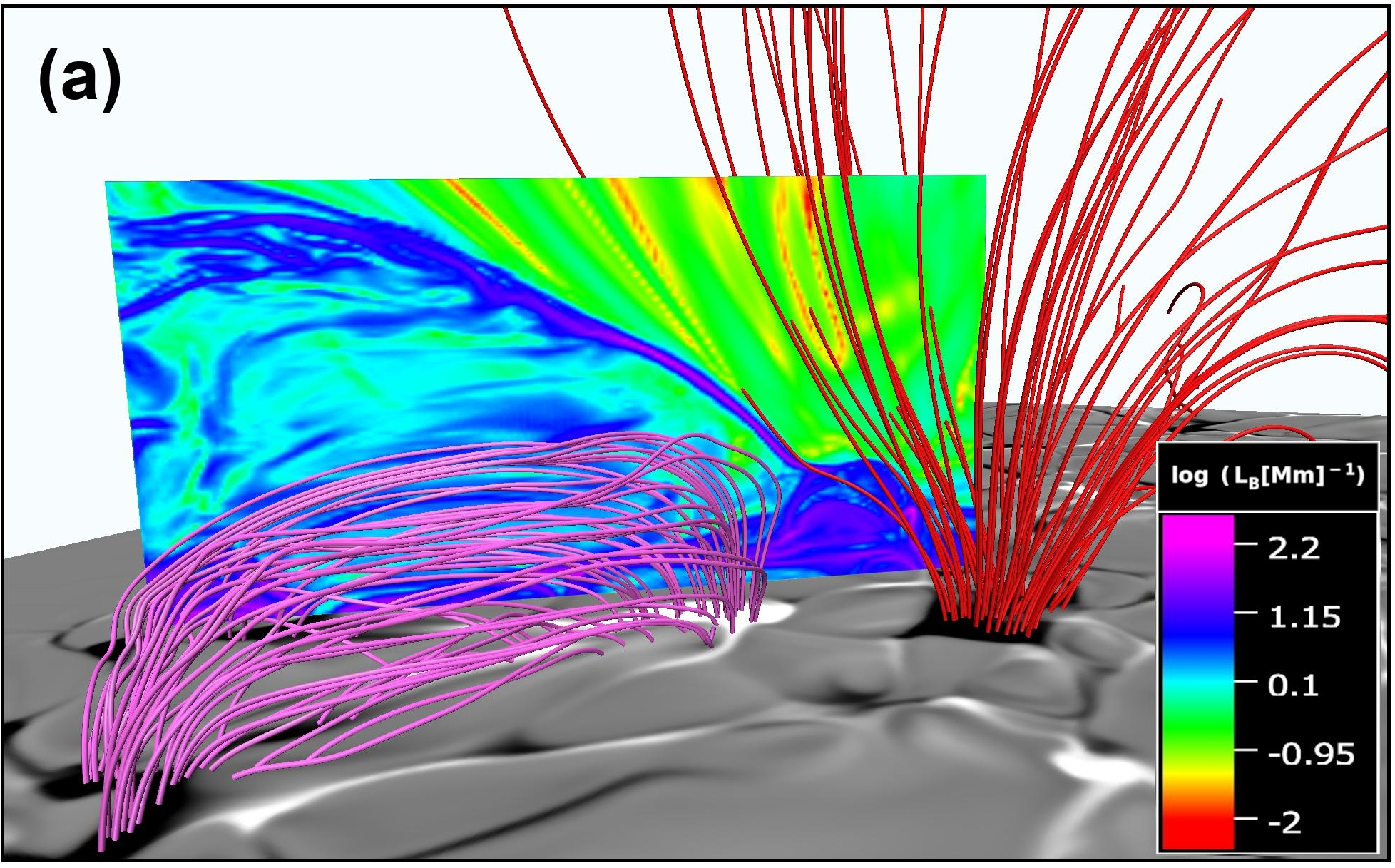}
\vskip 5mm
\hskip 0.05mm
\includegraphics[width=0.45\textwidth]{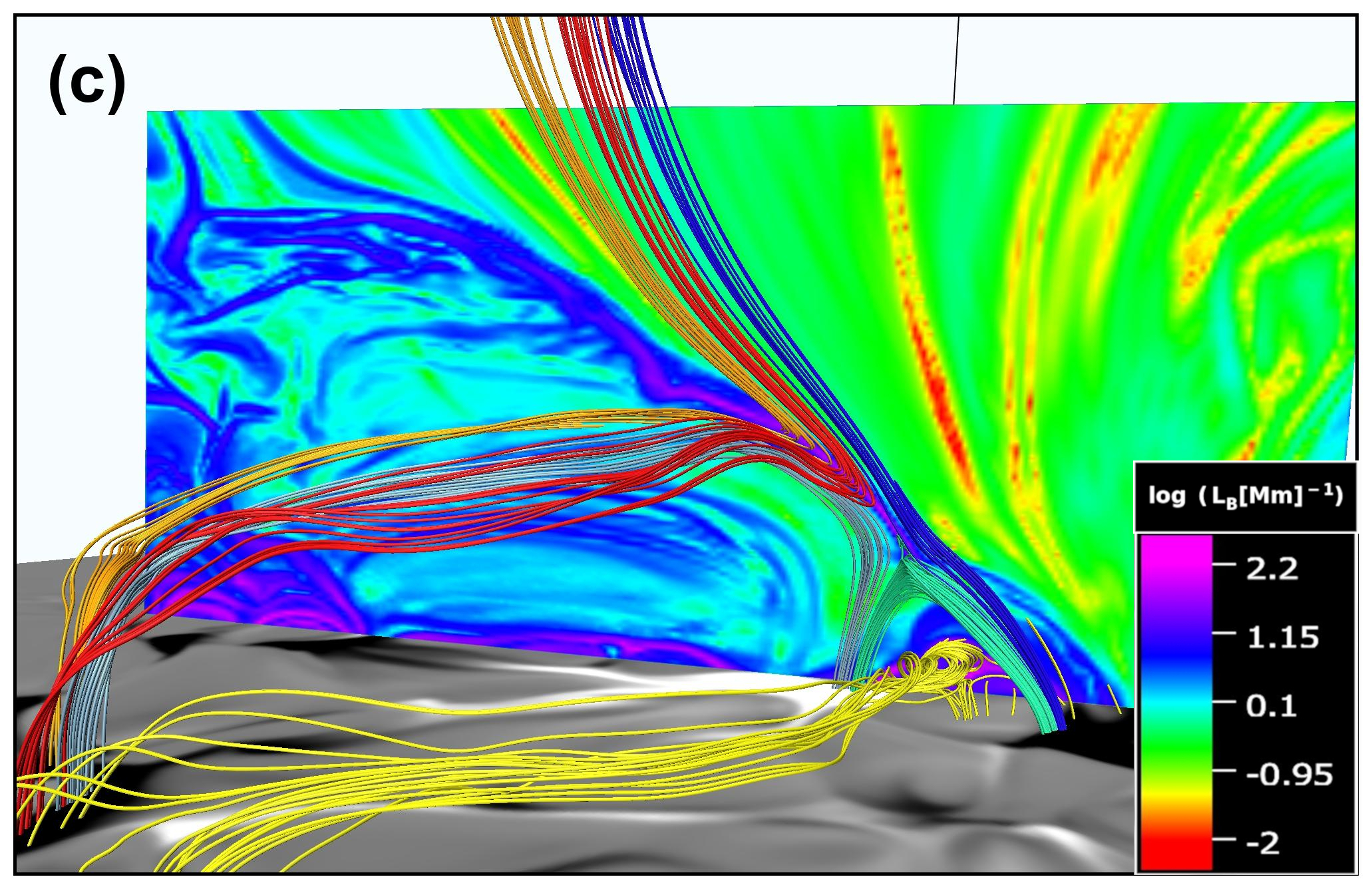}
\vskip 0.6mm}
\hskip 2.5mm
\includegraphics[width=0.530\textwidth]{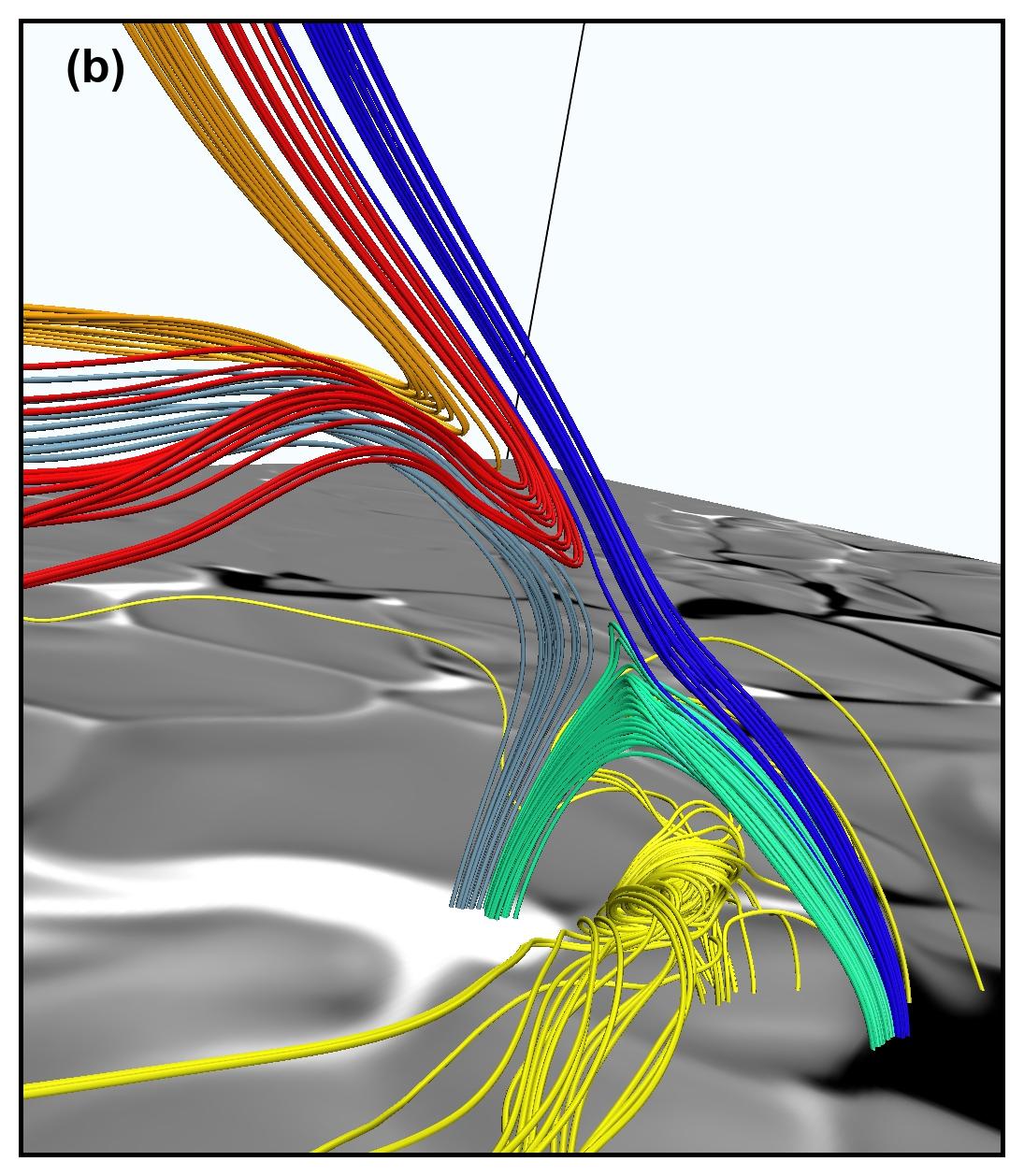}
}
\vskip 3mm
\centerline{
\includegraphics[width=0.346\textwidth]{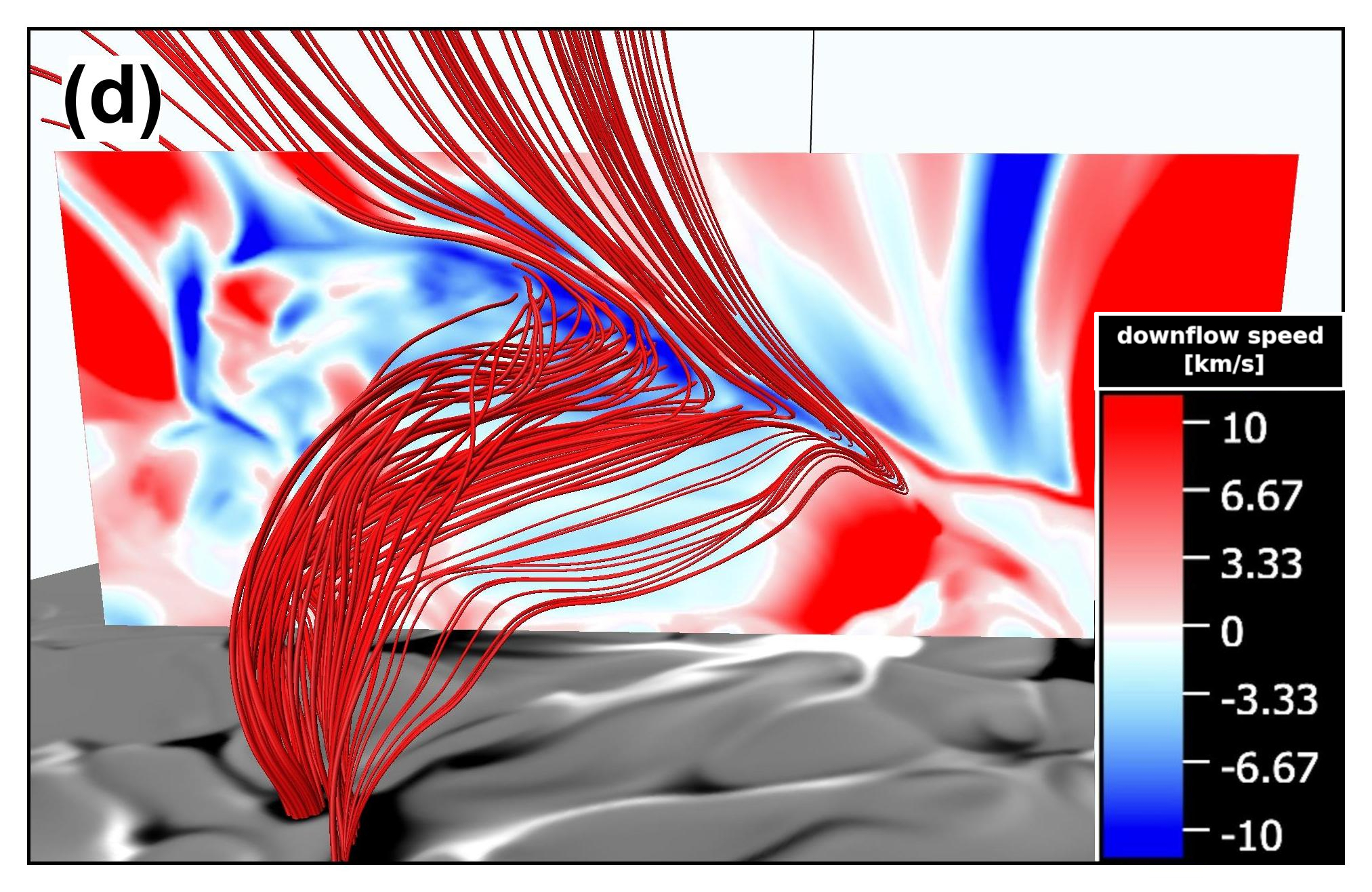}
\vbox{\hsize 0.34\textwidth
\includegraphics[width=0.34\textwidth]{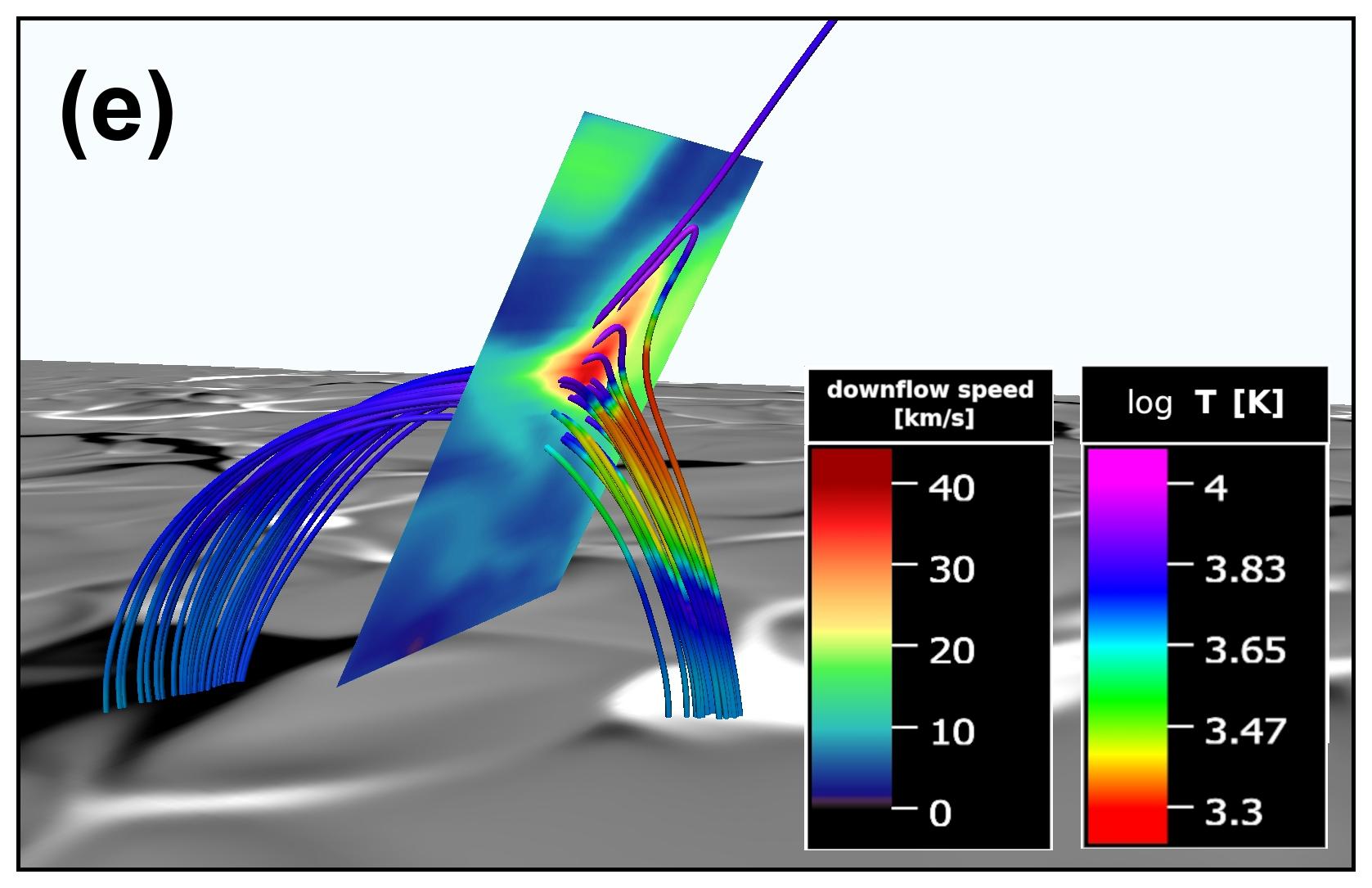}
\vskip 0.5mm}
\includegraphics[width=0.305\textwidth]{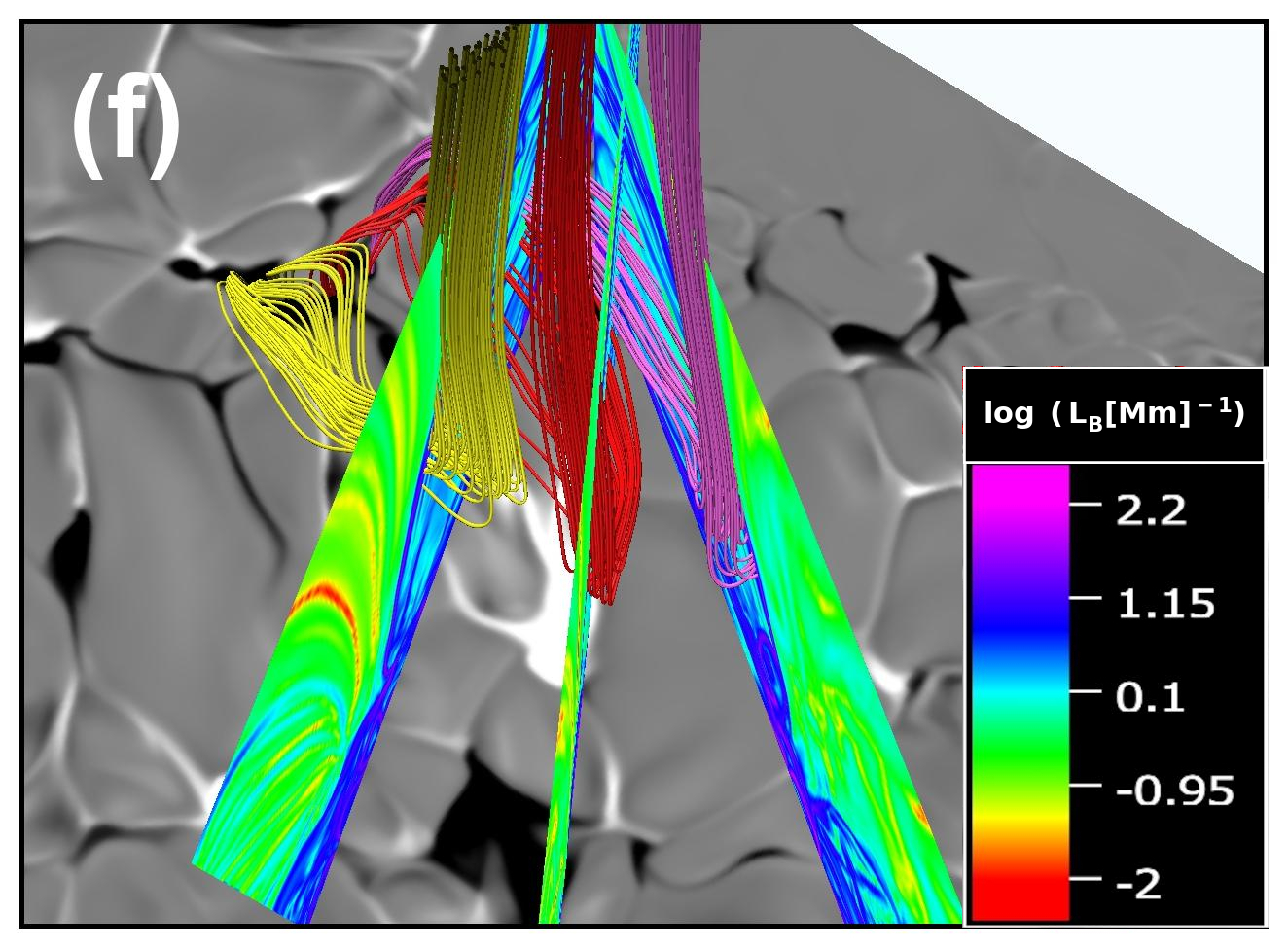}
}
\caption{3D visualization of the general field line configuration during Phase 1 using a snapshot at $t=113.3$~min.  Top left (panel a): Closed loops attached to the emerging bipole (purple) and field lines from the preexisting negative polarity that link to the general coronal field in the simulation (red).  Panels (b) and (c): six field line sets illustrating the reconnection process, with different colors for each set, as explained in the text.  The vertical plane in panels (a) and (c) subtends an angle of $42$~deg to the $XZ$ plane and contains a colormap for $\log(\invLB)$, with $L_B$ in~Mm.  Panels (d) and (e): detail of the upgoing and downgoing reconected field systems, respectively. The vertical plane contains a color map of the downflow speed in~\kms. The field lines in panel (e) are colored according to $\log T$. (f): Three recently reconnected field lines sets with vertices pointing to a common reconnection site in the region above the main photospheric magnetic patches. Colormap and -bar are as in panels (a) and (c).
\label{fig:flux_emergence_phase}
}
\end{figure*}

\section{Results: The 3D structure}\label{sec:3D_view}

In this section, we study the 3D configuration of the magnetic field, flows and thermodynamic variables at various stages of the cancellation process.  The major flux emergence episode leads to a conflict of magnetic orientations in the upper chromosphere long before any hint of cancellation at photospheric or low chromospheric levels is apparent. The interface between the emerged and preexisting magnetic systems becomes a thin current sheet across which a spatially abrupt change of connectivity of the field lines can be ascertained. This current sheet is therefore a blanket-shaped separatrix or quasi-separatrix layer, favorable to reconnection.

To illustrate the magnetic configuration above the colliding polarities we select two distinct phases. Phase~$1$ starts at about $t\approx 108$~min (left column in Fig.~\ref{fig:cancellation_site_1_general}). At this time, in the photosphere (say, in $z \lesssim 0.5$~Mm), the emerging field already shows a clear bipolar appearance; the approaching positive polarity of the bipole is still separated by about $2$~Mm from the main preexisting negative polarity at those heights. In fact, in this phase, those polarities are on a collision course, but they are separated by a weakly magnetized granular cell of shrinking size as the polarities get closer to each other.  We set the end of this phase at about $t\approx 115 - 116$~min, when the gap (or strait) between the colliding polarities is nearly closed and the photospheric observational cancellation phase has started. Checking with the flux profiles of the lower panel in Fig.~\ref{fig:flux_patches}, we see that these two phases coincide with the ascending and descending sections, respectively, of the positive polarity flux curve in it (blue).

\subsection{The first phase of the 3D cancellation process}
\label{sec:3D_structure_and_evolution}

\subsubsection{The reconnected field line systems}\label{sec:reconnected_field_line_systems}

Figure~\ref{fig:flux_emergence_phase}, taken at $t=113.3$~min, exhibits a typical configuration during the first phase.  Panel (a) shows two sets of field lines illustrating the simple linkage (i)~within the emerging bipole (purple) and (ii)~from the preexisting negative polarity to the general field in the corona (red). To provide a visual impression of the current sheet, a vertical map for the inverse characteristic length of the magnetic field variation, defined through:
\begin{equation}\label{eq:inverse_length}
\invLB \;=: \; \frac{|\nabla \times \Bvec|}
{|\Bvec|}\;=\;\frac{|\mu_0\,\jvec|}{|\Bvec|}\,,
\end{equation}
has been added. This map clearly shows that a current sheet has been created through the contact of the emerging and preexisting polarities; that current sheet envelops the emerged domain, and reaches down to a roughly triangular region peaking at about $z=0.8-1.0$~Mm at that time. The field configuration in the current sheet and in the triangular region can be discerned through panel (c) and the blow-up thereof (panel b). We see that reconnection is in full swing in the heights above the {\it strait} separating the colliding polarities. Additionally, the field lines in yellow show that a twisted flux rope has been created below the closed loops and above the intervening granule.

Focusing on the reconnection process, the colliding field line systems are not mutually antiparallel, since they have unrelated sources, and 3D reconnection ensues. The reconnection yields both closed post-reconnection loops and field lines that link to the general coronal field that existed prior to the emergence: the red and orange field lines in panel (b) in Fig.~\ref{fig:flux_emergence_phase} correspond to the latter, i.e., to hybrid field lines that now link the emerged bipole to the general coronal field; they have strong-curvature segments, apparent as sharp V-shapes in the region where they are leaving the reconnection site; they link the negative polarity of the emerging bipole to the coronal field. The field lines in green are closed post-reconnection loops that bridge the gap between the opposite photospheric patches; these field lines are warped at the top, i.e., they also have large-curvature segments when leaving the reconnection site.  The height of the reconnection site at the time shown in Fig.~\ref{fig:flux_emergence_phase} varies as one moves in the longitudinal direction along the separating strait between the colliding polarities, but it can be estimated to be between $1$ and $1.5$~Mm. We note that there is no null point associated with this reconnection: the minimum field strength in and around the current sheet is some $20$~G; the current sheet corresponds to a quasi-separatrix layer (QSL) across which reconnection is taking place.

The downflow/upflow configuration can be seen in panel (d) of Fig.~\ref{fig:flux_emergence_phase}, where a colormap of the downflow speed is plotted together with an extended collection of reconnected field lines on the upflow side: upwards-oriented velocities with $z$-component of size about $10$~\kms (blue colors) are detected all along the upper branch of the current sheet; the Lorentz force is dominant in it and drives the plasma upward along the QSL, with the field lines gradually losing the sharp curvature they had upon leaving the reconnection site.  Taking into account the horizontal component of the velocity, the total speed on the upflow side reaches values up to $20$~\kms. On the other hand, from the chromospheric diagnostics discussed in Sect.~\ref{sec:chromospheric_brightenings_and_jets}, we expect important downflows, like $20$~\kms, on the other side of the reconnection site. This is indeed the case, as apparent in panel (e) in the same figure: the closed post-reconnection loops, are seen to be ejected with large downflow speeds of up to $30$~\kms, which are in fact, super-magnetosonic. The highly warped shape of the loops is gradually flattened out. The high-downflow region ends abruptly at a fast-mode shock; the post-reconnection region is thus reminiscent of the equivalent region /downward reconnection outflow in the classical CSHKP flare model \citep[see, e.g.,][]{Priest_Forbes_2002, Shibata_Magara_2011}. Further details on the link between the MHD structure of the region and the observational proxies is provided in Sect.~\ref{sec:the_caII_jet}.

Even though the two flux systems have a conflict of orientations across the whole current sheet separating them, most of the reconnection seems to be taking place preferentially above the narrow region separating the magnetic concentrations. The reconnection site, therefore, has a limited extent in the horizontal direction; neighboring reconnected field line sets leaving the reconnection site run upward along the current sheet, {\it fanning out} horizontally, so to say, from the place where they suffered the reconnection. This can be glimpsed by comparing the three field line sets (in yellow, red and purple) in panel (f) of Fig.~\ref{fig:flux_emergence_phase}. All of them have sharp V-shapes but their vertices point toward a common reconnection region above the two colliding photospheric polarities.

\subsubsection{The current sheet and quasi-separatrix layer: average
  quantities and kinetic energy estimate}\label{sec:current_sheet_estimates} 

\newcommand{\textinbox}[3]{\vbox to 8mm{\hsize #1 \vfill\vskip 2pt
\centerline{#2}\vskip 4pt \centerline{#3}\vfill}}

\begin{table*}[tbp]
\begin{center}
\begin{tabular}{|c|c|c|c|c|c|c|c|c|}
\hline
     \textinbox{1.6cm}{$<|\vvec|>$}{[\,\kms\,]}
   & \textinbox{1.6cm}{$<u_z>$}{[\,\kms]\,}
   & \textinbox{1.6cm}{$<T>$}{[\,K\,]$^{\vbox to 2mm{\vfill}}$}
   & \textinbox{1.6cm}{$<\rho>$}{[\,g~cm$^{-3}$\,]}
   & \textinbox{1.4cm}{$<B>$}{[\,G\,]$^{\vbox to 2mm{\vfill}}$}
   & \textinbox{1.4cm}{$<z>$}{[\,Mm\,]$^{\vbox to 2.5mm{\vfill}}$}
   & \textinbox{1.8cm}{$<\rho\,v^2/2>$}{[\,erg~cm$^{-3}$\,]}
   & \textinbox{2.8cm}{total kinetic energy}{[\,erg\,]}\\
\hline
 $6.5$ & $5.5$  & $\pot{6.4}{3}$ & $\pot{1.2}{-10}$ & $30$ &  $1.8$ & $22.0$ & $\pot{6.7}{24}$ \\
\hline
\end{tabular}
\caption{Average values of different quantities in the current sheet above the emerging bipole.}\label{table:averages_in_current_sheet}
\end{center}
\end{table*}

\begin{figure}[htbp]
\vbox {\hsize 0.4\textwidth 
\centerline{\includegraphics[width=0.36\textwidth]{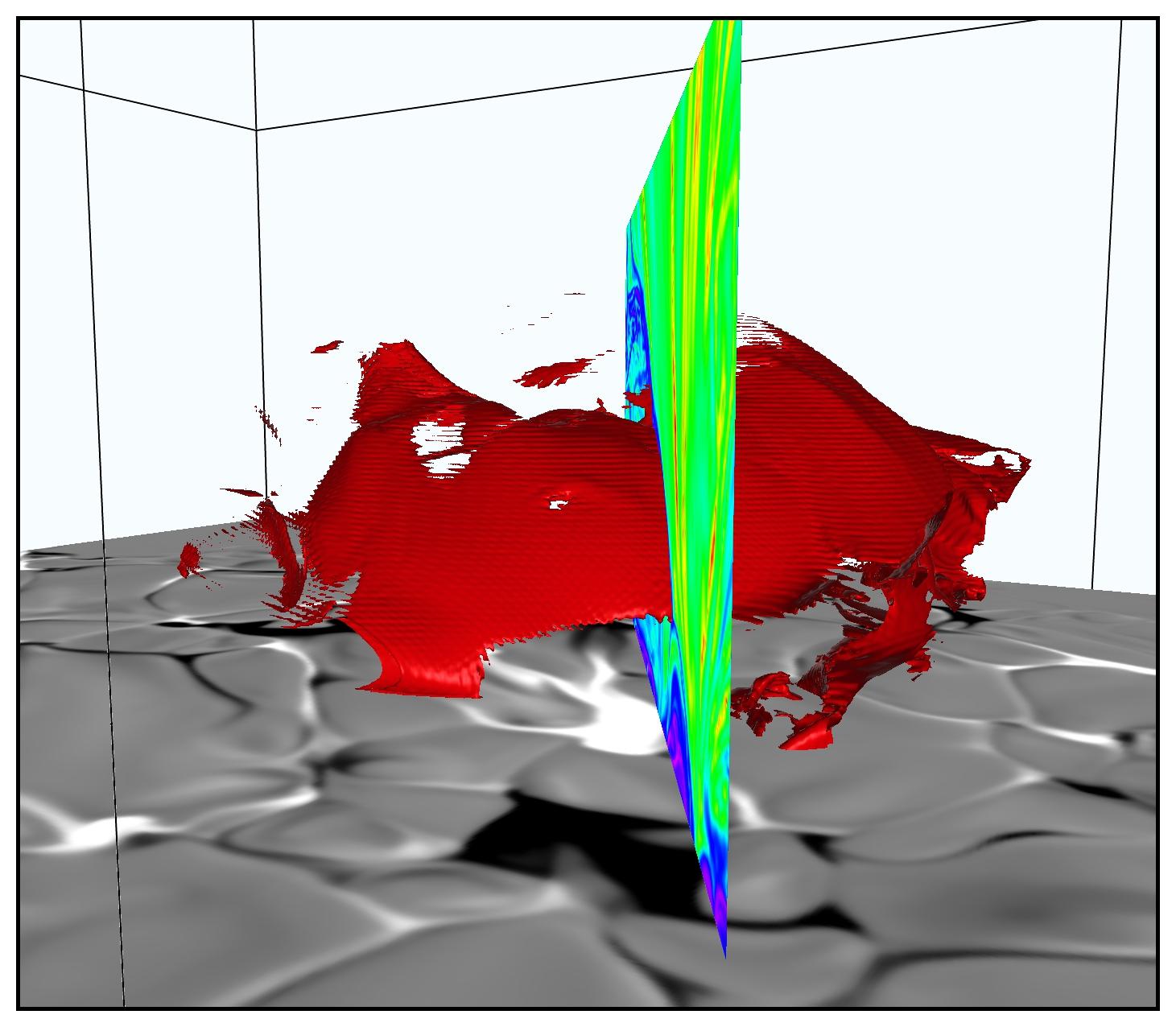}}
\vskip -2mm
\centerline{\includegraphics[width=0.45\textwidth]{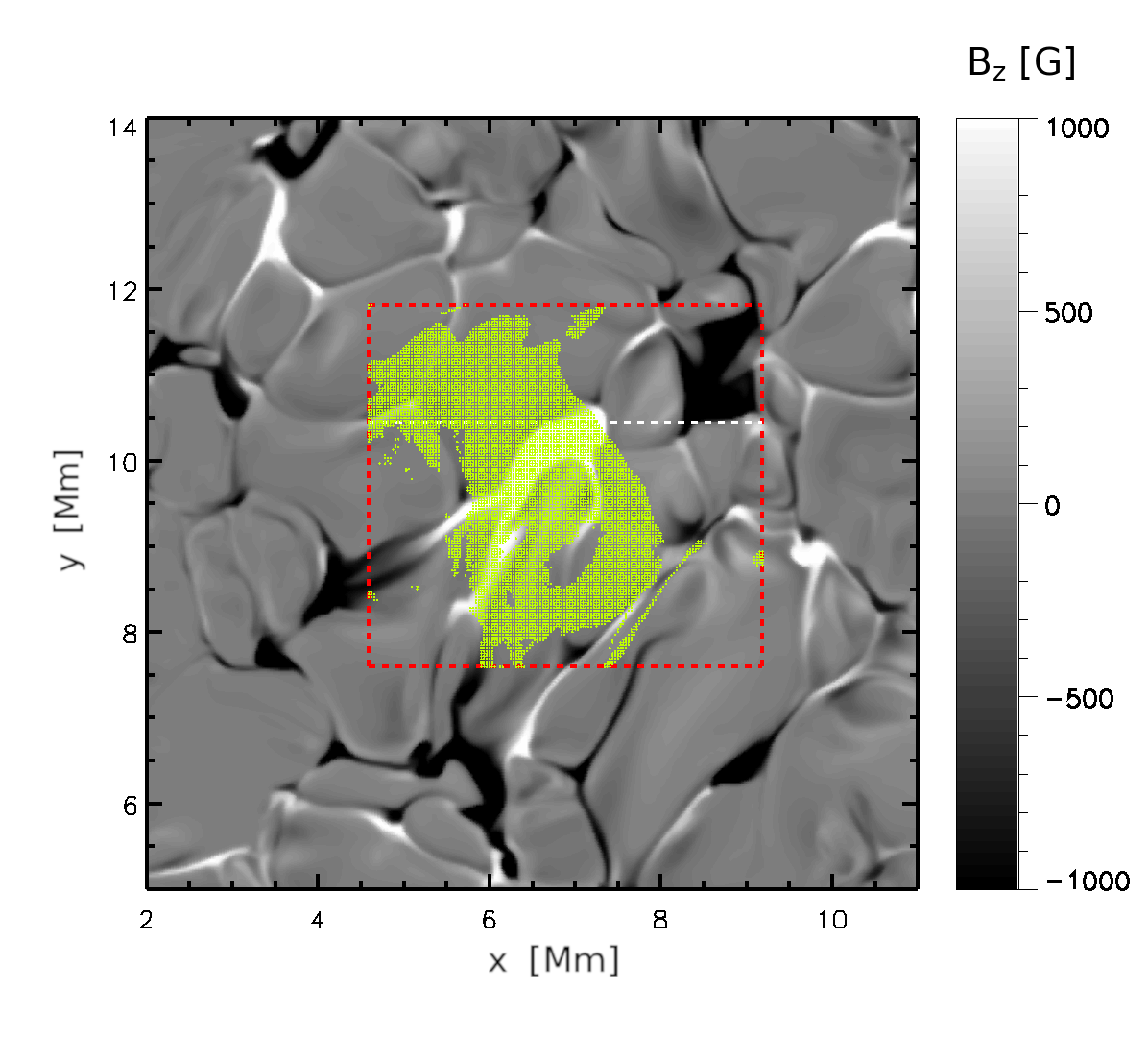}}
\vskip -2mm
\centerline{\includegraphics[width=0.45\textwidth]{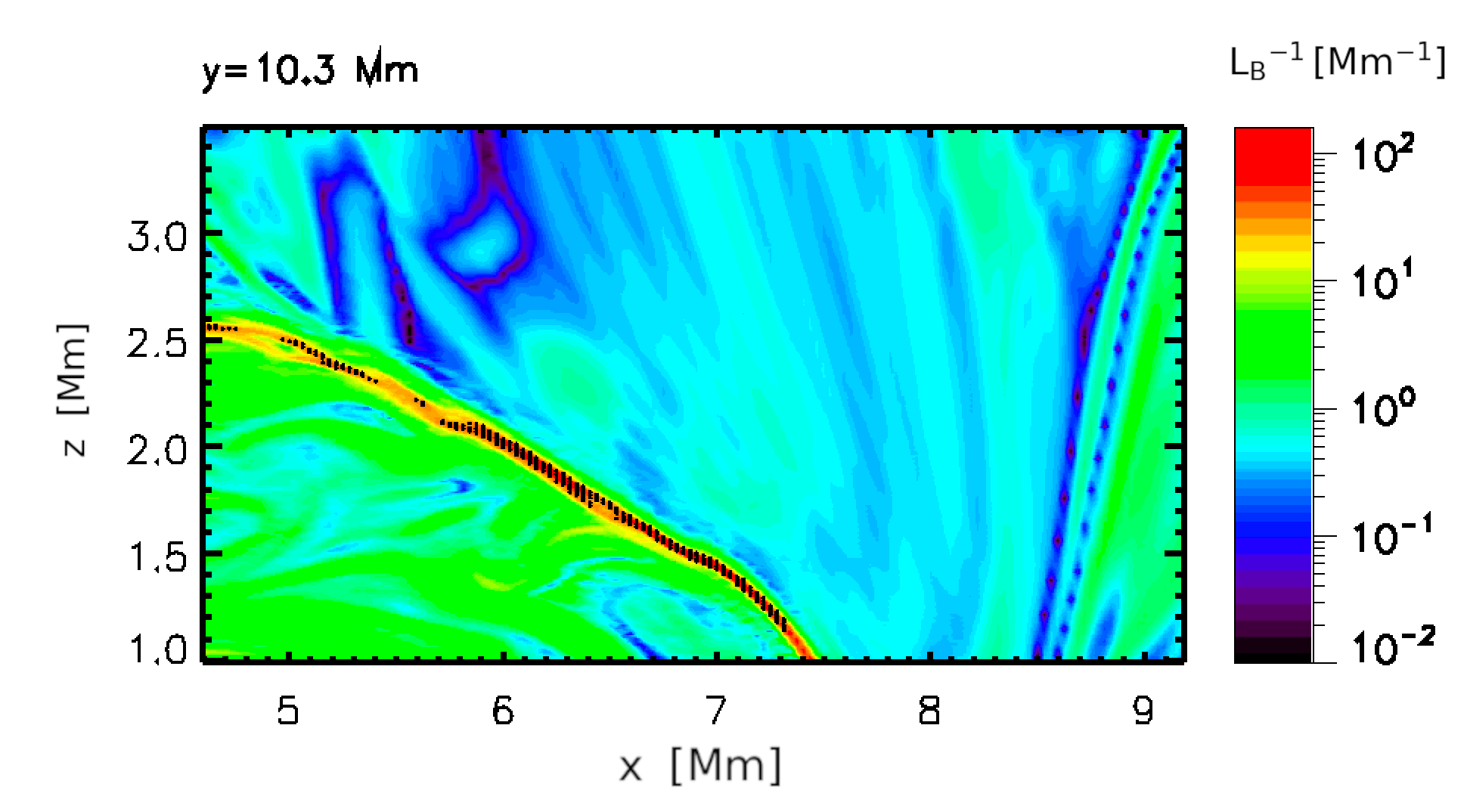}}
}
\caption{Statistics on the current sheet at $t=113.3$~min.  Top: Isosurface $\log(\invLBinMm) = 1.4$ in the region above the colliding polarities. The vertical map is the same as in Fig.~\ref{fig:flux_emergence_phase}, panels (a) or (c); the horizontal background is the magnetogram at $z=0$.  Center: Top view of the region selected for calculation of average quantities in the current sheet. The base of the cuboid where the high-$\invLB$-test was carried out is delineated as a red-dashed rectangle; the $(x,y)$-locations of the pixels fulfilling the criterion are marked in green, all of them restricted to the height range $(1, 3.5)$~Mm. Bottom: Color map of $\invLBinMm$ on a vertical cut at $y=10.3$~Mm with, superimposed, the points on the map with $\invLB > 10^{1.5}$~Mm$^{-1}$. The position of the vertical cut at $y=10.3$~Mm has been marked in the middle panel with a white dashed line.
}\label{fig:current_sheet_3D} 
\end{figure}

For a better understanding of the nature of the reconnection that is taking place in the current sheet separating the two canceling polarities, one can estimate average values of velocity, density, temperature or kinetic energy in the upward-moving outflow region. The current sheet is thin (as apparent in the vertical color maps for $\invLB$ in Fig.~\ref{fig:flux_emergence_phase}, panels a or c), and has the shape of a blanket covering the side of the emerged bipole facing the preexisting open-field polarity. This can be seen in Fig.~\ref{fig:current_sheet_3D}, top panel, which shows the isosurface $\log (\invLBinMm) = 1.4$ in red. The vertical plane in the figure is the same as that of panel (b) in Fig.~\ref{fig:flux_emergence_phase}.

To obtain rough estimates of relevant quantities in the current sheet, we consider, at $t=113.3$~min, a rectangular cuboid of horizontal size $4.6 \times 4.2$ Mm$^2$ that covers the two colliding photospheric polarities and a large part of the loops of the emerging bipole; in height, the cuboid is contained within the horizontal planes at $z=1$~and~$3.5$~Mm. The projection of the top and bottom lids of the cuboid can be seen in the middle panel of Fig.~\ref{fig:current_sheet_3D} as a rectangle in red on the background of a photospheric magnetogram. The pixels to be used in the statistics are those fulfilling $\invLB > 10^{1.5}$~Mm$^{-1}$ and with upward-pointing velocity (i.e., $u_z \ge 0$) within that volume.  The $(x,y)$ positions of those pixels are shown in the middle panel through green markers superimposed on the $z=0$ magnetogram. The vertical projection of the red isosurface of the top panel roughly coincides with the green area in the middle panel; we also see that the $\invLB$-isosurface covers the positive side of the emerged bipole, but extends a substantial distance sideways from the main photospheric field concentration. Finally, a vertical cut at $y=10.3$~Mm containing a color map of $\invLBinMm$ is shown in the bottom panel, with, superimposed in black, the points fulfilling the minimum-$\invLB$ criterion.

The resulting averages for the module of the velocity,~$|\vvec|$, upflow speed~$u_z$, temperature, density, field strength and height of those pixels are as given in Table~\ref{table:averages_in_current_sheet}.  The values of average density, temperature and height match typical mid-chromospheric values in 1D semi-empirical models, like between $0.9$ and $1.5$~Mm in the VALC model \citep{VALIIIC_1981}. The value of $<u_z>$ is well below the high values found close to the reconnection site in the previous section, which can be taken as an indication that the plasma elements are decelerated as they run upwards along the current sheet.  In the final two columns, the table also includes the average kinetic energy per unit volume and the aggregated kinetic energy in the selected points.  There is a large scatter of the values for all those variables in the current sheet. Yet, in spite of the rough nature of these estimates, we can conclude that the reconnection process going on above the cancellation site at this stage is chromospheric reconnection.

\subsection{Twisted flux ropes arising during the cancellation process}\label{sec:twisted_flux_ropes} 

As secondary consequences of the dynamical and magnetic evolution associated with the expansion of the emerging bipole and/or with the reconnection in the cancellation site, we have detected two coherent twisted flux ropes.

\begin{figure}[htbp]
\centering
\centerline{\includegraphics[width=0.50\textwidth]{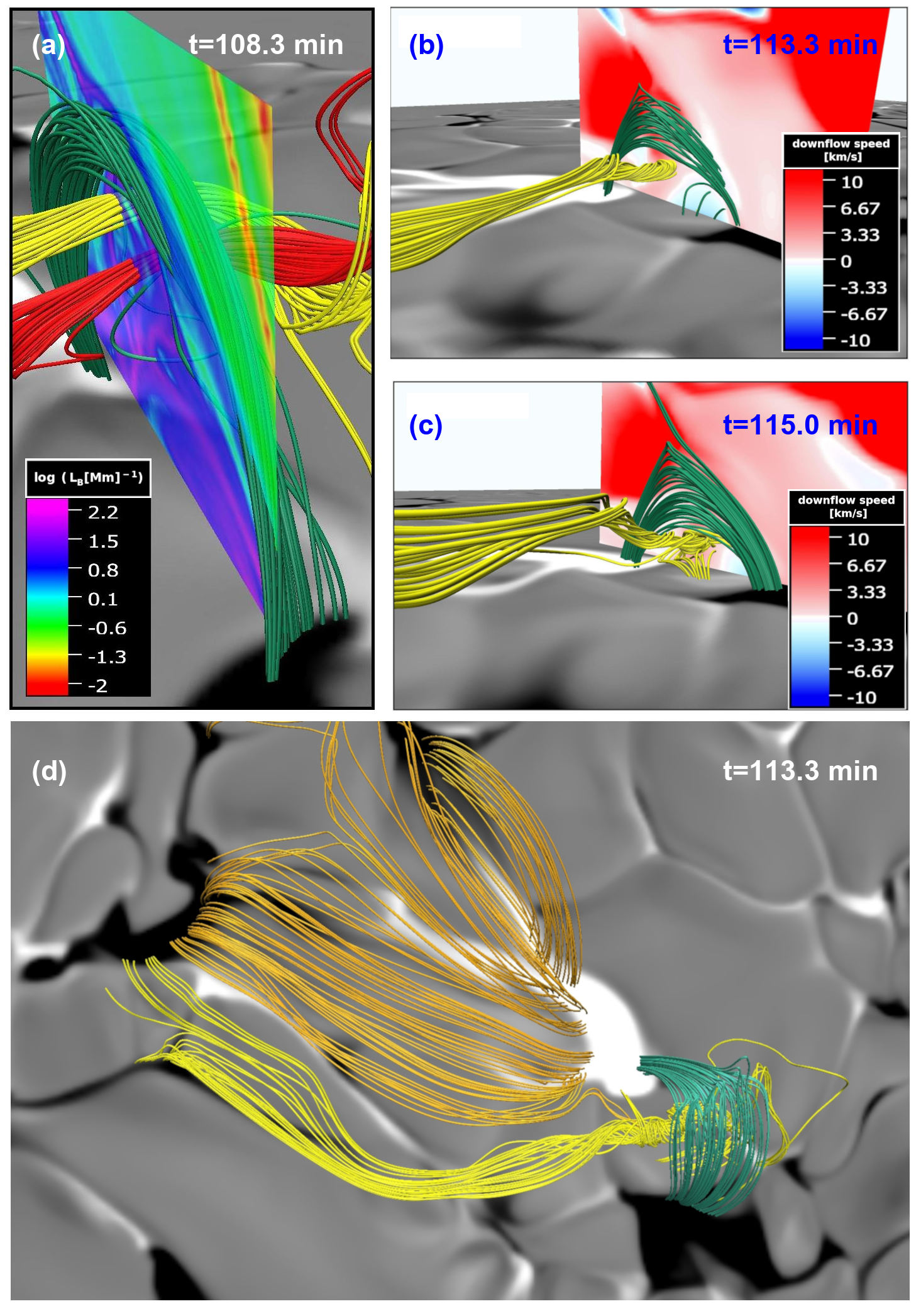}}
\caption{The twisted flux rope below the post-reconnection loops. (a) :~early stage ($t=108.3$~min), with braided flux strands (like the yellow and red ones) below the post-reconnection loops (in green) at heights $500$-$700$~km above the photosphere. The vertical map is for $\log(\invLBinMm)$.  (b):~intermediate stage ($t=113.3$~min), with a twisted flux rope (in yellow) below the post-reconnection loops (in green); the vertical map is the downflow speed, in \kms. (c):~final stage, at $t=115.0$~min. (d):~view of the general emerged field at $t=113.3$~min (orange), next to field lines similar to those of panel (b) (in yellow in both panels).
\label{fig:fluxrope_1}}
\end{figure}

\begin{figure}[bp]
\centerline{\includegraphics[width=0.45\textwidth]{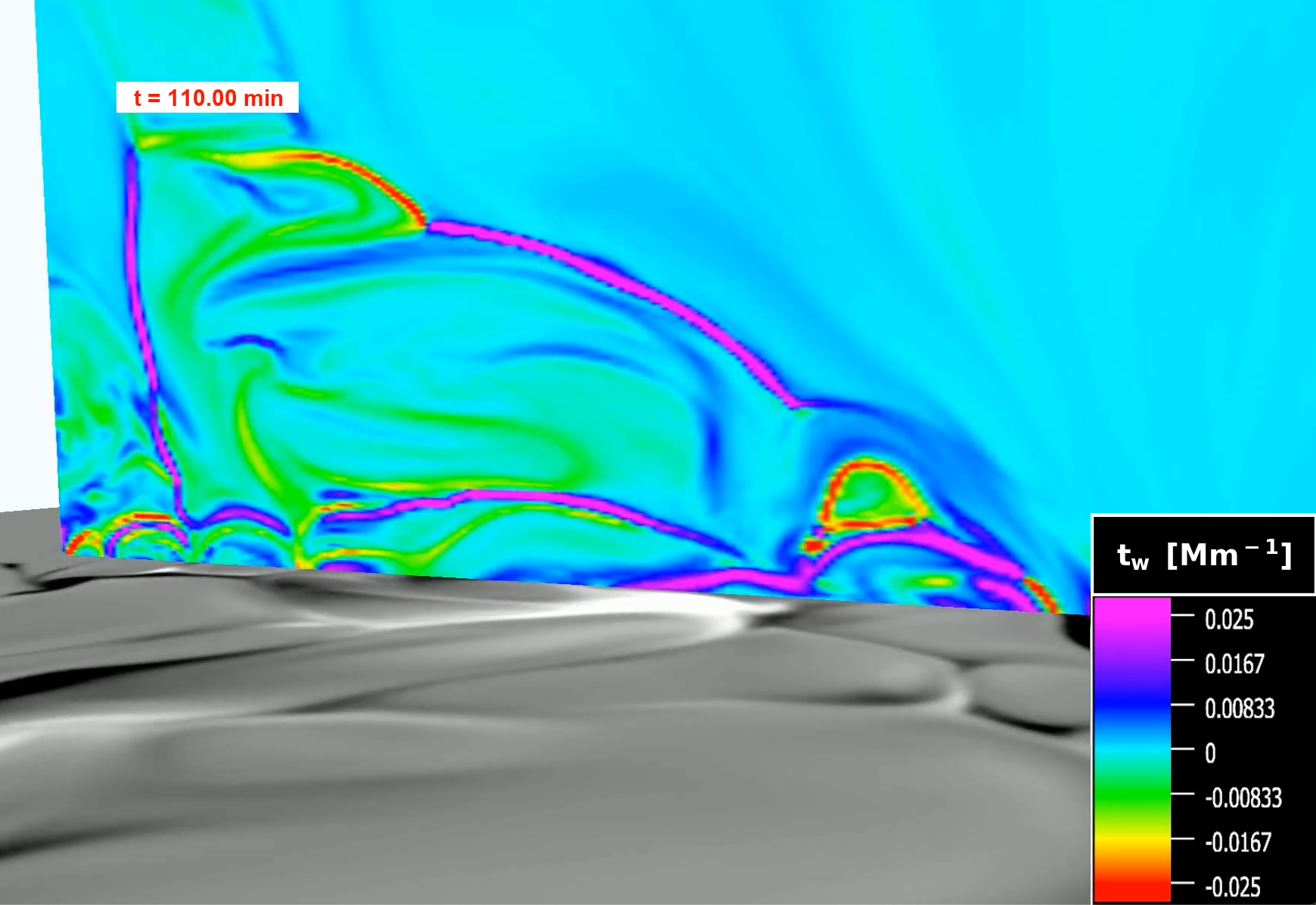}}
\caption{Distribution of the twist density (Eq.~\ref{eq:twist_parameter}) on a vertical plane. The ring of negative values in red delineates the instantaneous location of the twisted flux rope. The associated animation shows the time evolution of this figure between $t=108.3$~min and $t=116.7$~min.  (An associated animation is available for this figure.)  }
 \label{fig:twist}
\end{figure}

\subsubsection{The twisted flux rope underneath the post-reconnection loops
} \label{sec:flux_rope}

Looking back at Fig.~\ref{fig:flux_emergence_phase} we can see that the post-reconnection loops in the first phase of the cancellation process do not reach all the way down to the photosphere; rather, in the lowest several hundred km a twisted magnetic flux rope has formed, as delineated by the yellow field lines. This flux rope appears early on in the process of mutual approach of the canceling polarities. For instance, at $t=108.7$~min (corresponding to the magnetograms in the left column in Fig.~1), in the region above the strait separating the polarities one can see a collection of twisted field lines with negative current helicity, perhaps resulting from a slow vortical motion below the post-reconnection loops. Figure~\ref{fig:fluxrope_1}, panel (a), illustrates the field line configuration in this early stage (for $t=108.3$~min): two separate twisted threads (in yellow and red) appear braided around each other at heights roughly between $z=600$ and $800$~km above the photosphere; a strand of the post-reconnection loops is also shown in green.  Clearly, the granular flows cannot reach so high (the inverse granulation can extend up to, roughly, $300$~km above the $\tau_{500}=1$ level, see \citealt{Cheung_etal_inverse_granulation_2007}). In fact, through the vertical map for the electric current in the same panel we see the two topologically distinct regions underneath the reconnected loops: the top one has a roundish shape and harbors the incipient flux rope; below, one sees a small region, of some $200$~km in height directly perturbed by the underlying granule. The field lines in the latter (not shown in the figure) are flat arches linking opposite sides of the granule.  Some $5$~minutes later, at $t=113.3$~min (Fig.~\ref{fig:fluxrope_1}, panel b), the post-reconnection loops fill a larger fraction of the chamber below the reconnection site; there is now a well delineated twisted flux rope, located above the granule separating the polarities, some $300$~km above the photosphere. Through the color maps for the vertical velocity included in that panel, we can see that the twisted flux rope is being pushed downward with a speed of some $3$~\kms; we also see that next to the rope there is an upflow domain created by the underlying granule.  Some $\sim 2$~minutes later ($t=115.0$~min, Fig.~\ref{fig:fluxrope_1}, panel c), with the opposite polarities ever closer and thus nearer to the observational photospheric cancellation phase (Sect.~\ref{sec:main_cancellation_phase}), the post-reconnection loops occupy almost the whole domain below the reconnection region down to $100$~km above the photosphere, as illustrated through the green field line set; the downflowing region in which the flux rope was embedded is still present and has a speed of some $2$-$3$~\kms right down to the photosphere. The magnetic region that constituted the flux rope earlier on is thus reaching the photosphere, possibly being submerged to lower levels: the field lines in the near-photospheric levels are complicated and the twisted flux rope structure is more difficult to discern at that stage. When contact between the opposite-polarity patches is established at the photosphere, there is no further trace of a twisted flux rope between them.

The origin of this flux rope can be further glimpsed through Fig.~\ref{fig:fluxrope_1}, panels (d), (b) and (c). The emerged bipole has, at its flank, in high-photospheric or low-chromospheric heights, roughly horizontal field lines (like those in yellow in those panels).  When the reconnection starts at the advancing front of the bipole, some of these field lines are caught underneath the reconnection site. They are then slowly wound up, thus becoming the observed flux rope. The latter is then pushed down, as explained above, while the other segments of those field lines remain roughly horizontal and maintain their height.

An alternative view of the process of formation and descent of the flux rope can be gained through the animation associated with Fig.~\ref{fig:twist}.  The vertical map in it corresponds to the twist density, $t_w$, defined as
\begin{equation}\label{eq:twist_parameter}
t_w =: \frac{\mu_0}{4\,\pi}\,\frac{\jvec \cdot\Bvec}{B^2}\;.
\end{equation}
whose integral along any field line is the so-called twist number \citep{Berger_Prior_2006}. We note that $4\, \pi\, |t_w|$ is equal to $L_{B}^{-1}$ multiplied by the absolute value of the cosine of the angle subtended by $\jvec$ and $\Bvec$. $4\, \pi\, |t_w|$ is exactly $L_{B}^{-1}$ in force-free situations only.  In the animation, the included vertical map for the twist density delineates, in the early snapshots (e.g., at $t=108.3$~min), a roundish shape centered at $z=0.75$~Mm and with diameter roughly $0.6$~Mm. As time advances, this shape becomes a distinct ring of negative $t_w$-values, with decreasing size: that ring coincides with the intersection of the flux rope with the vertical plane and delineates the boundaries between topologically distinct regions. The ring systematically descends toward the photosphere (with its center down to, e.g., $z=0.56$~km at $t=110$~min, with diameter of $0.37$~Mm; $z=0.44$~km at $t=111.7$~min, diameter $0.20$~km; or $z=0.25$~km at $t=113.3$~min, with tiny radius of some $0.15$~km) until it is swallowed by the photospheric convection at around $t=115$~min.

\subsubsection{A second twisted flux rope in the periphery of the emerging
  domain}  \label{sec:the_second_flux_rope} 

\begin{figure}[htbp]
\includegraphics[width=0.49\textwidth]{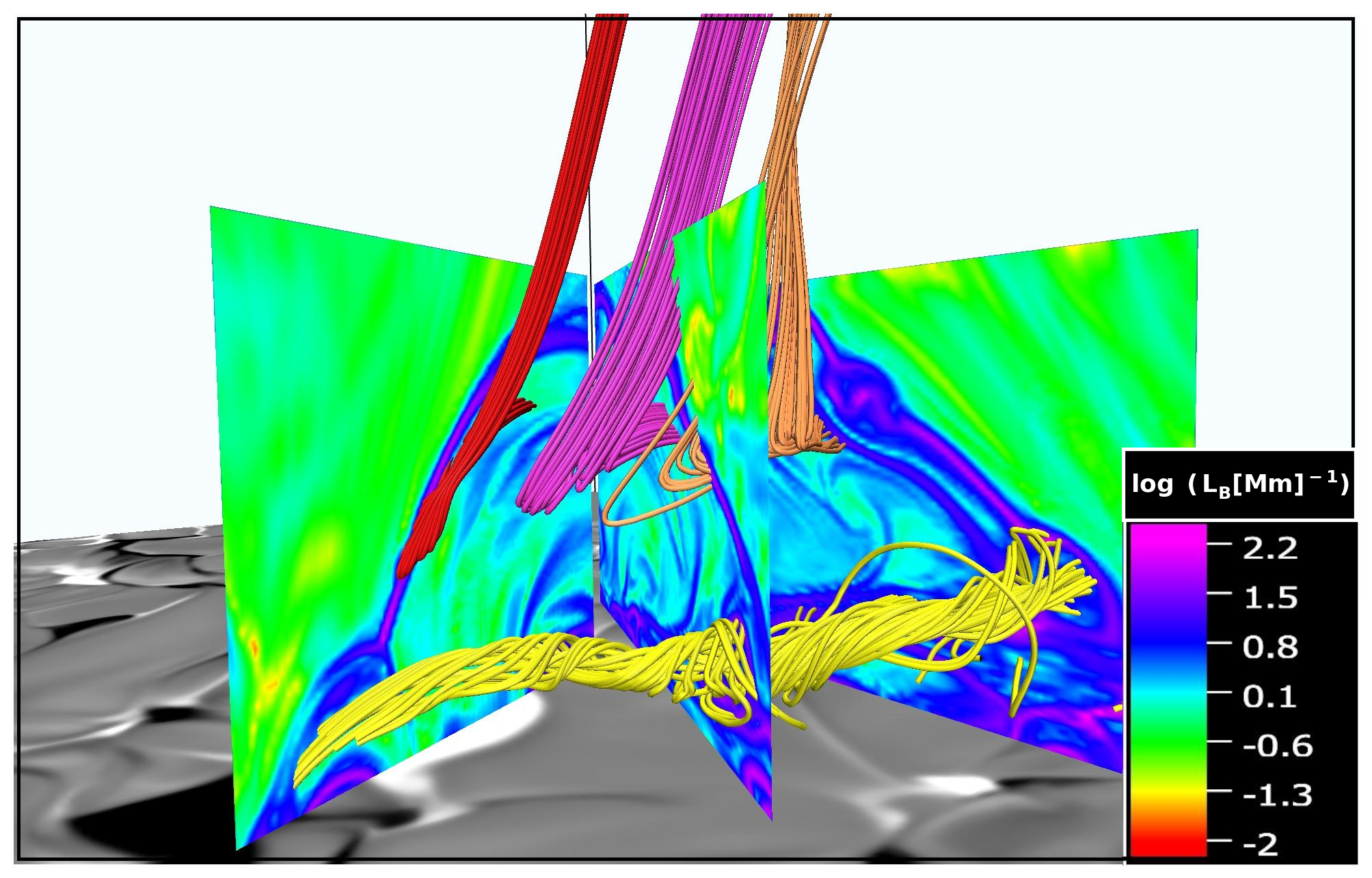}
\caption{View at $t=116.0$~min of the twisted magnetic flux rope arising at the front of the expanding emerging domain (yellow).  Three recently reconnected, V-shaped field line sets are also shown colored in red, purple and orange. The color maps on the vertical planes are for $\log(\invLBinMm)$.} \label{fig:long_flux_rope}
\end{figure}

We have identified another twisted flux rope, of considerable length, which extends well beyond the location where the main polarities are canceling. Figure~\ref{fig:long_flux_rope} shows various sets of field lines traced at time $t=116.0$~min and three vertical planes containing maps of $\log L_{B}^{-1}$. The canceling polarities can be seen on the magnetogram in the bottom-left of the figure; a collection of field lines, drawn in yellow, end in the negative polarity and delineate a twisted flux rope of negative current helicity. The main part of the rope is centered at a height of $0.7$-$0.9$~Mm. Its length between the shown vertical planes is close to $4$~Mm; the rope, therefore, is coherently created in a region that extends well away from the canceling polarities; this conclusion is also reinforced through the appearance of round-shaped current distributions in the vertical planes that delineate a volume of separate connectivity coinciding with the flux rope. The figure also contains three sets of V-shaped field lines (in red, purple and orange) similar to what was shown in panel (f) of Fig.~\ref{fig:flux_emergence_phase}.  It is difficult to ascertain the precise chain of events leading to the formation of this flux rope. Still, studying the time evolution of the interface at the side of the emerging bipole shown in Fig.~\ref{fig:long_flux_rope}, one concludes that the emerged domain, in its sideways expansion, pushes away the surrounding material; the roughly horizontal field lines at the indicated heights are rolled up in this process and lead to the coherent flux rope documented in this section.

\begin{figure}[bp]
\vtop{
\hbox to 8cm{\hfill\includegraphics[width=0.48\textwidth]{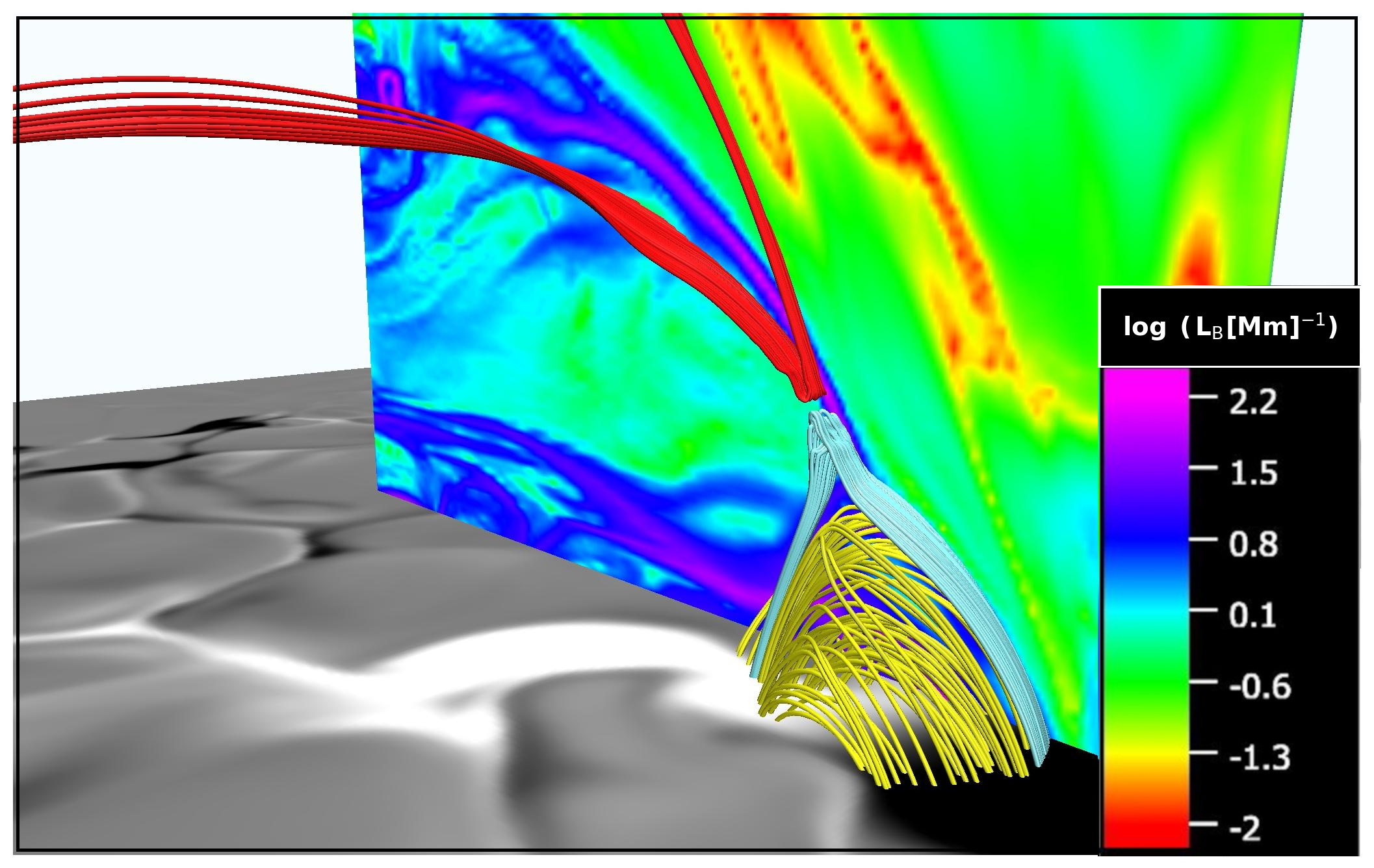}\hfill}
\vskip 1mm
\hbox to 8cm{\hfill
\includegraphics[
width=0.48\textwidth]{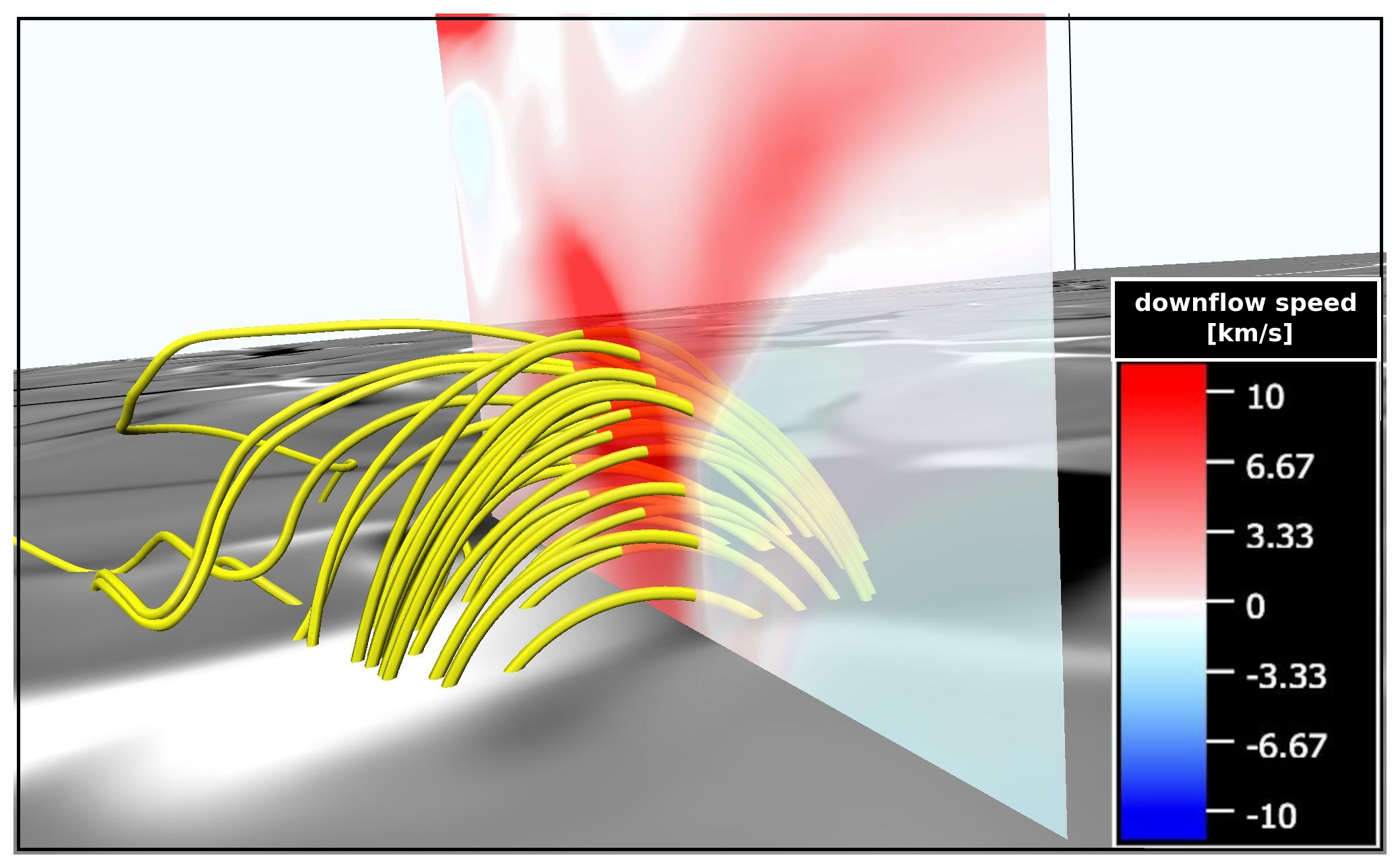}\hfill}
\vfill}
\caption{Selected field lines at $t=117.0$~min illustrating the reconnection process in the photospheric cancellation phase. Upper panel: fieldlines for three magnetic systems. Red: just-reconnected V-shaped upward-running fieldlines; light-blue: just-reconnected closed loops; yellow: closed post-reconnection loops filling the space below the reconnection point down to the photosphere.  Vertical plane: color map of $\log(\invLBinMm)$. Lower panel: color map for the downflow speed (in \kms) on a plane that cuts across the post-reconnection domain together with simple post-reconnection loops.}
 \label{fig:reconnection_main_cancellation_phase}
\end{figure}

\subsection{Photospheric cancellation stage: contact of the polarities at the photosphere and submergence of the post-reconnection loops.}
\label{sec:main_cancellation_phase}

All along the evolution during the first phase described in the previous subsections, the advancing opposite polarities gradually close the {\it strait} or {\it gap} separating them at the photosphere. By $t\approx 115$~min we can say that {\it the observational photospheric cancellation phase} has begun. Reconnection continues taking place at chromospheric levels, but now the domain occupied by the post-reconnection loops reaches all the way down to the photosphere. This can be seen in Fig.~\ref{fig:reconnection_main_cancellation_phase} (upper panel), which shows field lines associated with the reconnection process at $t=117.0$~min, so in the initial stages of this main photospheric cancellation phase; the red and light-blue field lines are recently-reconnected field lines and have high-curvature segments, with associated large Lorentz force. The reconnection site lies between those fieldline sets, roughly at $z=0.95\pm 0.1$~Mm, which is not very different to the height we found in the early reconnection phase. The yellow field lines are downward-moving post-reconnection loops which have lost much of the high curvature of the light-blue fieldlines. In contrast to the situation in Sect.~\ref{sec:reconnected_field_line_systems}, this fieldline system reaches the photosphere, which is the characteristic feature of this photospheric cancellation phase. Additional information can be gained from the lower panel: drawing a colormap for the downflow on a vertical plane that runs along the dividing region between the polarities, one sees that the post-reconnection loops are moving downward at large speeds, up to $25$~\kms when expelled from the reconnection site, down to about $5$~\kms when reaching the photosphere. From the photosphere, the loops further retract to subphotospheric levels.  Checking back with Fig.~\ref{fig:flux_patches} we see that both the positive and negative polarities are losing flux at comparable rates starting at around $t=114$~min all the way to at least $t=120$~min.  The reconnection site on the vertical plane shown in the upper panel (and neighboring ones) can be located at about $z=1$~Mm, although the precise location of the diffusion region cannot be uniquely determined given the complicated 3D geometry.

The configuration depicted in Fig.~\ref{fig:reconnection_main_cancellation_phase} is maintained in its general traits all along this main photospheric cancellation phase, at least up to $t=124.0$~min. A little later, a negative polarity coming from another flux emergence process taking place at greater $y$ reaches the site and distorts the cancellation process, so we do not discuss the 3D structure at those later times.

\subsection{The 3D plasma and magnetic field configuration leading to 
salient chromospheric spectral features} \label{sec:the_caII_jet} 

The vigorous reconnection occurring at the cancellation site leads to observational consequences already partially described in Sect.~\ref{sec:chromospheric_brightenings_and_jets}. Using the 3D study of the foregoing subsections, we can now attempt to link salient features in the \CaIIIR~spectra or filtergrams with localized regions of high velocity and/or density in the 3D volume near the reconnection~site.

\begin{figure}[t]
\centerline{\includegraphics[width=0.45\textwidth]{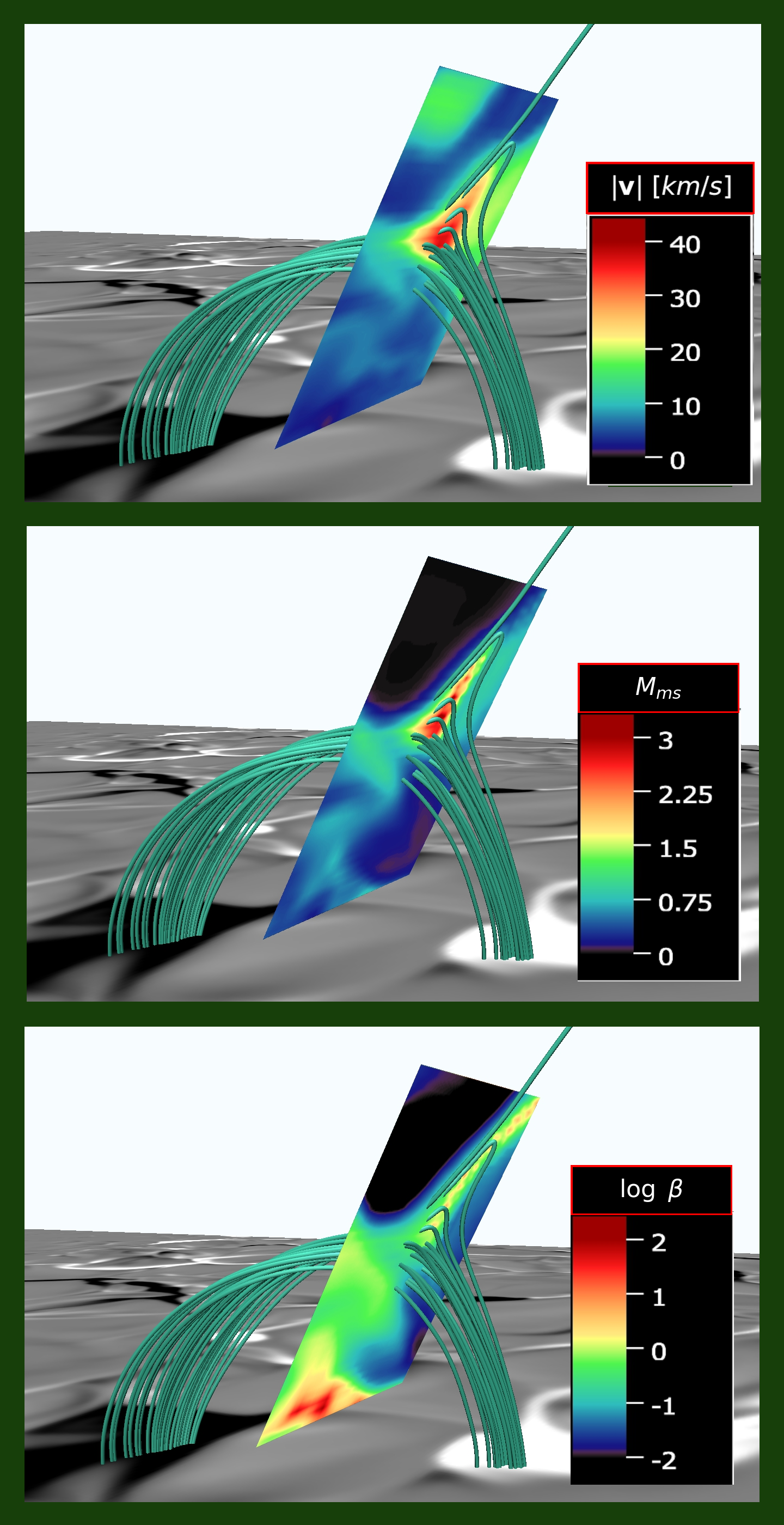}}
\caption{The region of supercritical downflow originating in the reconnection site. The three panels show a collection of closed reconnected loops; the colormap is (a) for the module of the velocity in \kms (top); (b) the magnetosonic Mach number (center), and (c) the logarithm of the plasma $\beta$ (bottom).
 \label{fig:the_postreconnection_shock}
}
\end{figure}

Perhaps the most obvious feature in the spectra is the bright ``spike'' found in the red wing of all the Ca~II and Mg~II lines, centered in the location of maximal reconnection, e.g. as seen in the right column of Fig.~\ref{fig:spec_site1_ca8542_caK_mgk}. We can confidently assign this feature to the newly reconnected field lines that eventually form closed post-reconnection loops, shown in green in Fig.~\ref{fig:flux_emergence_phase} (panels b, c or e).  These field lines experience large velocities near the reconnection site before slowing as they straighten.  The initial plasma velocities reach up to $\approx 35$~\kms, which are both supersonic and super-Alfv\'enic, given that temperatures in the reconnection site are $\lesssim 10^4$~K, and that the Alfven speed is of order $10$~\kms.  Figure~\ref{fig:the_postreconnection_shock} focuses on the region directly after the reconnection. The top panel shows that the velocity $|\vvec|$ indeed reaches up to $35$~\kms. The high-$|\vvec|$ region has a sharp cutoff exactly where the post-reconnection loops are no longer highly warped.  Viewing the colormap for the magnetosonic Mach number, $M_{ms}$, (central panel), we see that the downflows directly resulting from the reconnection are super-magnetosonic, with $4> M_{ms}>1$ down to the sharp cutoff. So, the cutoff probably marks the site of a termination, fast-mode shock, where the plasma attached to the post-reconnection loops is braked in its downward motion and deflected in the longitudinal direction along the field lines. The panel on the right, finally, shows the logarithm of the plasma $\beta$: the values in the funnel-like high-speed post-reconnection region are within about a factor $2$ around unity. Even if the gas and magnetic pressures are not very different, the high field-line curvature surely makes the Lorentz force predominant compared with the gas pressure gradient. Based on the $10$-sec cadence synthetic spectra calculated, this downflow is relatively steady, and moves with the reconnecting region.

In contrast, the signature of the up- and sideways-accelerated gas coming from the upper portion of the reconnection region, shown by the red and orange field lines in Fig.~\ref{fig:flux_emergence_phase}, is more episodic. Furthermore, since the field lines there are closer to horizontally oriented, the motion does not cause as large a Doppler shift in the line profiles. Even so, the motion shifts the highest opacity of the calcium and magnesium lines $> 20$~\kms to the blue and this high velocity gas appears as much brighter than the surroundings at this wavelength. This is because the surroundings are formed several hundred kilometers below in a similar fashion as seen for the red wing in the third column of Fig.~\ref{fig:spec_site1_ca8542_caK_mgk}, near the temperature minimum region of the chromosphere, where temperatures are significantly lower. When viewed at a spectral frequency in the blue part of the \CaIIIR, \CaIIK\ or \MgIIk\ lines equivalent to the highest velocities found, $\simeq 25$~\kms, emission from this outflow appears as a jet moving rapidly from the reconnection site, as seen in Fig.~\ref{fig:ca8542_caK_mgk3_flame}. At spectral frequencies closer to the line core, indicative of lower velocities, the fan shaped structure of the accelerated plasma, seen in Fig.~\ref{fig:flux_emergence_phase}(f) or Fig.~\ref{fig:long_flux_rope}, becomes more apparent and outlines essentially the entire current sheet described in Sect.~\ref{sec:reconnected_field_line_systems} and \ref{sec:current_sheet_estimates}. We note that apart from the high, indeed supercritical, velocities incurred by reconnection, the plasma densities do not show large differences from the surroundings and temperatures are raised, but not above $10^4$~K; reconnection mainly serves to accelerate gas, not directly heat it. While we have shown the most dramatic event in Fig.~\ref{fig:ca8542_caK_mgk3_flame}, continual though sporadic events occur during the entire cancellation process and we find high velocity jets at $t=114$~min and $t=116$~min as well as the jet occurring between $t=(121,123)$~min described here. The emitting plasma of these jets is dense, and is driven mainly in a horizontal direction, thus never achieving TR nor coronal heights or temperatures and therefore remaining invisible to diagnostics formed at temperatures above some $10\,000$~K. On the other hand, in the spectral diagnostics we do not find any obvious trace of the flux ropes presented in Sect.~\ref{sec:flux_rope}.

\section{Summary and discussion}\label{sec:discussion_and_conclusions}

\noindent \textbf{5.1 Summary.}\nl 
In this paper we have analyzed a flux cancellation process occurring in a Bifrost numerical model stretching from the top layers of the convection zone to the corona. The cancellation is the result of a bipolar flux emergence event: in its expansion in the lower atmospheric layers, the bipole's positive polarity meets a preexisting negative polarity and cancellation follows. We have first studied a number of features often encountered in observational cancellation studies: magnetograms, total unsigned magnetic flux, horizontal field strength and vertical velocity at the PIL, synthetic maps and spectra for the \CaIIIR, \CaIIK, and \MgIIk\ lines.  We have then explored the 3D properties of the reconnection between the canceling flux systems. The reconnection occurs at chromospheric heights, has no null points, and yields two sets of reconnected field lines: (a) V-shaped magnetic field lines that run toward the corona along the blanket-like current sheet covering the emerged bipole; (b) downgoing reconnected field lines that have strongly warped shapes which are soon ironed out by the Lorentz force. The reconnection outflows, which reach supersonic and super-Alfv\'enic speeds of up to $30$~\kms, have chromospheric densities and temperatures, and lead to observable brightenings and spikes in the red or blue wing of the chromospheric spectral lines, both in synthetic filtergrams and in the spectra themselves. The apparent cancellation seen in magnetograms first starts in the chromosphere while the photospheric polarities are still well separated by a granular cell. In advanced stages, the canceling polarities finally meet at the photosphere, but the reconnection is still chromospheric; the post-reconnection loops run downwards all the way to the surface and submerge into subphotospheric depths. Twisted flux ropes are seen to form in different parts of the canceling structure. 

\noindent \textbf{5.2 RMHD modeling from the convection zone to the corona and the cancellation events.} \nl 
Having 3D data from a radiation-MHD model that includes the photosphere, chromosphere and corona allows us a detailed and encompassing view of the cancellation process impossible to obtain through observational means alone.  In observations, cancellation is often understood as the progressive disappearance of equal amounts of positive and negative vertical magnetic flux upon contact of two opposite-polarity patches in the photosphere (or, less frequently, in the chromosphere). However, the phenomenology of cancellation involves the corona as well, with, for instance, coronal bright points often being associated with cancellation sites \citep[e.g.][]{Harvey_etal_1999, Madjarska_2019}. In the case analyzed in this paper, the main reconnection site is located in chromospheric layers, with correspondingly low-temperature and high-density outflows, but the inclusion of coronal heights up to $14.5$~Mm above the photosphere permits us to follow the upward-going reconnected field lines without interference from the boundary conditions. Also, the march of the opposite polarities toward each other and the magnetic pattern in low photospheric layers is modified by the intervening granulation. All of this, and the possibility of synthesizing observables, points to the importance of having a radiation-MHD treatment of the low atmospheric layers in the models and of the inclusion of coronal heights in the experiment.  On the other hand, in numerical simulations both with ideal-MHD and radiation-MHD codes, also in the current one, the plasma emerging from the photosphere reaches low temperatures as it expands into low pressure domains in the chromosphere in spite of the energy release via formation of the $H_2$~molecule \citep[see, e.g.,][] {Hansteen_etal_2019, Carlsson_etal_ARAA_2019}; low temperatures are also known to appear in chromospheric shock rarefactions \citep[e.g.][]{Leenaarts_etal_2011}.  To alleviate the problem, a minimum temperature, typically around $2$~or~$2.5 \times 10^3$~K, is prescribed (see, e.g., \citealt{Gudiksen_etal_2011} and \citealt{Przybylski_etal_2022}, for the Bifrost and MuRAM codes, respectively).  While there is observational evidence of reduced temperatures in emerging bubbles \citep{Ortiz_etal_2014, De_la_Cruz_etal_2015, Nobrega-Siverio_etal_2024}, the detected reduction is less extreme than in the experiments.  Although detailed analysis of the effects of nonequilibrium ionization and partial ionization has been carried out in flux emergence experiments \citep{Nobrega-Siverio_etal_2020}, it is still unclear which physical processes can operate in the Sun that may possibly prevent this temperature decrease \citep[see discussion in][]{Carlsson_etal_ARAA_2019, Martinez-Sykora_etal_2023, Evans_etal_2025}.

\noindent \textbf{5.3 Flux emergence, cancellation, coronal jetting activity.}\nl 
The emergence of magnetic bipoles from the solar interior can naturally lead to reconnection in the overlying magnetized atmosphere, with or without ensuing photospheric cancellation. This has been shown in idealized numerical models along the past decades, with the upwards-directed reconnection outflows evolving into hot, smooth, collimated ejections in the corona \citep[e.g][]{Yokoyama_Shibata_1996, Archontis_etal_2004, Galsgaard_etal_2005, Moreno-Insertis_etal_2008}, which sometimes may be below the instrumental threshold of detection (\citealt{Nobrega-Siverio_etal_2024, Nobrega-Siverio_etal_2025}; see also \citealt{Sterling_etal_2016}).  For cancellation to follow, a nearby preexisting magnetic element must be in the neighborhood of the opposite polarity of the emerging bipole, so that, after convergence, the two systems mutually cancel, at least in part.  This pattern has often been seen, e.g., in the framework of CBP evolution \citep{Madjarska_2019}, where, in fact, eruptions tend to occur associated with the flux cancellation stage rather than the flux emergence one \citep{Mou_etal_2018}. In the case discussed in this paper emergence leads to reconnection and the attendant dynamical and thermal phenomena at chromospheric level, but it does not lead to any obvious coronal jets or hot coronal loops detectable in the observational proxies: not every instance of flux emergence followed by cancellation necessarily leads to a coronal jet. In our follow-up paper, instead, we will be encountering a case in which the cancellation does indeed produce hot loops and ejections at the coronal level.

\noindent \textbf{5.4 Reproduction of observational features}\nl 
The values obtained in Sect.~\ref{sec:flux_measurements} for the magnetic flux of the individual colliding patches (about $\pot{1.3}{19}$~Mx at the beginning of the photospheric cancellation phase) and flux decay rate, namely $\pot{2.7}{19}$~Mx~h$^{-1}$ per polarity, are within the standard ranges found in the observational literature. For instance, \citet{Park_etal_2009} used magnetograms of canceling magnetic features (CMFs) taken by the SOHO/MDI and Hinode/SOT; for the latter, they produced values for twelve CMFs with flux between $\pot{2}{18}$ and $\pot{3}{20}$~Mx. The corresponding flux decay rates were between about $\pot{3}{18}$~Mx~h$^{-1}$ and $\pot{6}{19}$~Mx~h$^{-1}$. A number of other authors, e.g., \citet{Chae_etal_2004, Chae_etal_2010, Bellot_Beck_2005, Jiang_etal_2007, Gosic_etal_2018} or \citet{Kaithakkal_Solanki_2019}, provided flux cancellation data with flux and flux decay rate ranging between ($\pot{1}{17}$~Mx, $\pot{4}{17}$~Mx~h$^{-1}$) for the smallest elements \citep{Kaithakkal_Solanki_2019}, and ($\pot{2.5}{20}$~Mx, $\pot{2}{19}$~Mx~h$^{-1}$) for the largest ones \citep{Jiang_etal_2007}. To our knowledge there is so far no systematic study of the dependence of the flux decay rate with the total flux of the canceling elements nor with the type of solar domain considered (quiet Sun, coronal hole, plage, active region), but the numbers from our simulation seem to be compatible with those obtained in the observations.

For the horizontal field strength $\Bhor$ and vertical component of the velocity at the PIL we have obtained average values of $630$~G and $3.7$~\kms downflow, respectively, when calculated at a height of $z=0.15$~Mm, with a fairly good correlation between those two quantities; for the PIL at $z=1$~Mm, instead, the downflows, even if still marginally predominant, show no important correlation with $\Bhor$. The detection of significant values for $\langle \Bhor\rangle$ and vertical velocity at the PIL between canceling polarities has been considered in many observational papers in the past. The values of the horizontal field in our study are compatible with those found in observational papers that find significant transverse fields and downflows in the PIL like 
(a)~\citet{Chae_etal_2004}, who find $\langle \Bhor \rangle \sim 800$~G for a PIL in the low photosphere at a site in an active region;
(b)~\citet{Kubo_Shimizu_2007}, who observed twelve cancellation events in four active regions, in all of which the opposite polarities developed a horizontal-field linkage at photospheric heights as part of the cancellation process with $\langle \Bhor \rangle \sim 500$~G;
(c)~\citet{Chae_etal_2010}, who give $\langle \Bhor\rangle \sim 800$~G for a cancellation site in the quiet Sun;
(d)~all of them above those found by \citet{Iida_etal_2010}, who give $\langle \Bhor \rangle \sim 200$~G, also in the quiet Sun.
The downflow speed in their measurements tends to be smaller than in our case: $1$~\kms \citep{Chae_etal_2004, Chae_etal_2010}, $0.5 - 1$~\kms when near the disk center \citep[][their Fig.~10]{Kubo_Shimizu_2007}, and peak of $1.7$~\kms, average of $1$~\kms, for \citet{Iida_etal_2010}.  Still, further examples show that significant horizontal fields are not always detected at the PIL in observational studies, perhaps as a result of different magnetic configurations of the cancellation site, or, also, through differences in the formation height of the spectral line(s) used in the observation with respect to the height of the reconnection site.  For example, \citet{Bellot_Beck_2005}, through inversions of six photospheric spectral lines found no significant horizontal field in the PIL and upflows of about $1$~\kms in one of the two opposite polarities in a cancellation site at the outer edge of a sunspot moat.  \citet{Kubo_etal_2010b, Kubo_etal_2014} got negative results concerning the presence of horizontal fields linking the polarities in four out of five of their observed cases and argued that the lack of space and time resolution may prevent the detection of unequivocal cancellation signals. On the other hand, \citet{Fischer_2011} obtained horizontal fields at photospheric levels in $25$ out of the $33$ cancellation events she observed but could not come to conclusions on the occurrence of flux submergence or U-loop rise.  \citet{Kaithakkal_Solanki_2019} found transient horizontal fields in the PIL in their study of 11 photospheric cancellation events already mentioned in the text.

We find remarkable similarities in the chromospheric response to cancellation between our results and the observations by \citet{Gosic_etal_2018} despite important differences: internetwork field strengths and environments in their case versus the stronger field regions and more active flux emergence that characterize the present model. In particular this applies to the \CaIIIR~line, especially the time span of the interaction, the increased asymmetry in the line as compared with the surrounding environment (including the increased intensity of the near red wing around 0.02--0.04~nm from line center), and the intensity enhancement at all wavelengths. On the other hand, we also find clear indications of upflows in \CaIIIR\ as well as in the \MgIIk~and \CaIIK~lines along the current sheet shown in Sect~\ref{sec:current_sheet_estimates}, but not as strong and steady as the signal in the red, probably a result of fluctuations in the reconnection pattern.

While the asymmetry of the core peaks results from the velocities dominant in the canceling site, the core width of the \MgIIk\ line mirrors both density of material in the chromosphere and/or the turbulent velocities found in the line forming region \citep[see e.g.][]{Hansteen_etal_2023, Ondratscheck_etal_2024}.  Other simulations of the upper chromosphere have encountered difficulties in reproducing widths as observed by IRIS, even in models with relatively high resolution, such as the current one with $\Delta x$ of order 30~km \citep{Hansteen_etal_2023}. Parametric studies \citep{Carlsson_etal_2015} have shown that this problem can be alleviated by increasing the amount of mass in the chromosphere, or alternately with high amplitude turbulent motions. The latter has been demonstrated by \citet{Ondratscheck_etal_2024}. As can be seen in Fig.~\ref{fig:cancellation_site_1_line_profiles} we obtain line profiles that are not significantly narrower than what is observed when averaged over the entire computational domain, and this is to a lesser extent also true for the average profile surrounding the canceling region. While highly turbulent motions are present in the region surrounding the cancellation site, in the model presented here it is the additional mass brought up by flux emergence over a large portion of the computational domain that is the most important contributor to \MgIIk\ widths.

\noindent \textbf{5.5 Twisted flux ropes}\nl
We have found a few instances of roughly horizontal, twisted flux ropes in topologically distinct volumes in the stages preceding the actual photospheric cancellation phase. A first one is formed below the closed post-reconnection loops at low chromospheric heights; it is pushed down by them, has a length of $\sim 1$~Mm when traversing the low photosphere, and ends up possibly submerging into subphotospheric layers. Another twisted flux rope is detected at low chromospheric heights in the flanks of the blanket-like current sheet between the two flux systems, it is longer (up to roughly $4$~Mm), and disappears due to the complicated dynamics of that region without reaching the low photosphere. Twisted flux ropes have attracted significant attention in recent years, especially as possible precursors of large ejections \citep[see, e.g.][and references therein]{Amari_etal_2025}. The flux ropes here are not the result of tearing-mode instabilities in the current sheet nor are they generated by shear flows stressing an arcade-like magnetic configuration. The twisted ropes found here seem to be quite coherent along their length; they last for some $6$~to~$8$~min (the first one) and at least some $3$~to~$5$~min (the second one), and naturally result from the dynamics of the domain underneath the closed post-reconnection loops (in the first case) or by the dynamics at the front of the expanding domain occupied by the emerged bipole (the second one). Their formation is surely facilitated by the presence of horizontal field in the flanks of the emerged dome at high-photospheric or low-chromospheric heights. It is likely that they are a natural occurrence in cancellation sites that follow a bipolar flux emergence episode.  Various additional interesting aspects of the complicated dynamics and topology of these ropes must be left out of the present paper.

These twisted flux ropes are different to the minifilaments in the observational studies of, e.g., \citet{Panesar_etal_2016, Panesar_etal_2017_a,Panesar_etal_2018_a,Adams_etal_2018, McGlasson_etal_2019,Tiwari_etal_2019a, Moore_etal_2022}. Their filaments appear as darkenings in the EUV observations, so they must be located at coronal heights and contain excess density that absorbs the hot emission from lower levels. Also, they are destabilized by the cancellation occurring below them, and end up erupting, leading to a blowout jet detected in the EUV or X-Ray observations.  The flux ropes in our simulation do not have excess density and the main reconnection process is not taking place below them. Also, they are not ejected, but, rather, disappear as a result of the interaction with the low-lying photospheric flows.

\noindent\textbf{5.6 The TR and coronal response.}\nl 
Concerning the coronal response to the cancellation events in the low atmosphere, this cancellation site leaves no discernible traces in TR or coronal diagnostics, even though the corona is directly involved in the magnetic and dynamical cancellation processes. On the other hand, we find clear TR and coronal markers in other cancellation sites in the same simulation with similar chromospheric diagnostics. This will be discussed in a forthcoming publication.

In conclusion, the use of radiation-MHD models stretching from the uppermost convection zone to the corona opens the door to a global understanding of the process of magnetic flux cancellation that includes the magnetic, dynamic, and radiative linkage between those layers. This encompassing view can fundamentally enhance and complement the use of imaging and spectroscopic observations in photospheric, chromospheric, TR or coronal lines as well as previous theoretical studies that focus on individual aspects of the phenomenon. Magnetic flux cancellation is both a crucial process in solar physics and one which encompasses all those layers in an essential way. We hope that the theoretical study of a limited number of comparatively simple cases using radiation-MHD simulations can provide key insights into the physical nature of this phenomenon.

\begin{acknowledgements}
This research has been supported by the European Research Council through the Synergy Grant number 810218 (``The Whole Sun'', ERC-2018-SyG); by the Spanish Ministry of Science, Innovation and Universities through project PGC2018-095832-B-I00; and by the Research Council of Norway (RCN) through its Centres of Excellence scheme, project number 262622.  VHH was also supported by NASA contract NNG09FA40C (IRIS) and by NASA grants 80NSSC20K1272 and 80NSSC24K0258.  The use of UCAR's VAPOR software \citep{VAPOR_1,VAPOR_2} is gratefully acknowledged.  This work also benefited from discussions at the International Space Science Institute (ISSI) in Bern, through ISSI International Team project \#535 \textit{Unraveling surges: a joint perspective from numerical models, observations, and machine learning}.
\end{acknowledgements}

\vspace{1mm}
%
%
\bibliographystyle{apj} 

\end{document}